\documentclass[%
  conference,%
  10pt,%
]{IEEEtran}

\usepackage[T1]{fontenc}
\usepackage[utf8]{inputenc}
\usepackage{adjustbox}
\usepackage{amssymb} % Math symbols
\usepackage{amsmath} % Math stuff
\allowdisplaybreaks
\usepackage{subcaption}
\usepackage{cite}
\usepackage{hyperref}
\usepackage{defs}
\graphicspath{{./pictures}{./pictures/example1}{./pictures/refactorings/pdf}}
\usepackage{algorithm2e}
\usepackage{xcolor}
\usepackage{graphicx}
\usepackage{stfloats}
\usepackage{array, colortbl}
\usepackage{cellspace}
\usepackage{paralist}
\newcommand\myincludegraphics[2][]{ \raisebox{-0.5\height}{\includegraphics[#1]{#2}}}

\newcommand{\blockleft}{\begin{mbox}\sf\begin{tikz}[baseline=(X.base)]\node[draw=black!30,fill=black!3,semithick,rectangle,inner sep=1pt, minimum size=1em, outer sep=0pt, rounded corners=1pt] (X)}%
\newcommand{\blockright}{;\end{tikz}\normalfont\end{mbox}}%
\newcommand{\scratchblock}[1]{\blockleft{\small\textsf{#1}}\blockright}
\newcommand{\blocklefttable}{\begin{mbox}\sf\begin{tikz}[baseline=(X.base)]\node[draw=black!30,fill=black!3,semithick,rectangle,inner sep=1pt, minimum size=1em, outer sep=0pt, rounded corners=1pt] (X)}%
\newcommand{\blockrighttable}{;\end{tikz}\normalfont\end{mbox}}%
\newcommand{\scratchblocktable}[1]{\blocklefttable{\scriptsize\textsf{#1}}\blockrighttable}
\newcommand{\numtransformations}{26}

\newcommand{\transformationsmean}{10.074041192566897}
\newcommand{\transformationscount}{7218237}

\newcommand{\swappercent}{\SI{25.95603330841035}{\percent}\xspace}

\newcommand{\mergescriptspercent}{\SI{16.757055774145403}{\percent}\xspace}

\newcommand{\splitslicepercent}{\SI{16.357041199949517}{\percent}\xspace}

\newcommand{\splitscriptpercent}{\SI{16.473454667670236}{\percent}\xspace}
\newcommand{\ififelsetoconjunctioncount}{3572\xspace}

\newcommand{\ififnottoifelsecount}{3572\xspace}

\newcommand{\conjunctiontoifelsecount}{3509\xspace}

\newcommand{\meanScriptCountNonRefactorable}{12.404}
\newcommand{\meanScriptCountRefactorable}{14.696706477151562}

\newcommand{\projectsNotRefactored}{28}
\newcommand{\projectsRefactored}{972}
\newcommand{\percentageProjectsDominated}{\SI{8.114}{\percent}\xspace}
\newcommand{\percentageRefactorableProjectsImprovedComplexityEntropy}{\SI{77.27}{\percent}\xspace}

\newcommand{\blocksImproved}{119}
\newcommand{\complexityImproved}{765}
\newcommand{\entropyImproved}{779}
\newcommand{\blocksWorsened}{742}
\newcommand{\complexityWorsened}{94}
\newcommand{\entropyWorsened}{82}
\newcommand{\blocksAvgEffect}{0.7325089133082981}
\newcommand{\complexityAvgEffect}{0.21632474540355612}
\newcommand{\entropyAvgEffect}{0.214709901512024}
\newcommand{\blocksComplexityImproved}{88}
\newcommand{\blocksEntropyImproved}{86}
\newcommand{\complexityEntropyImproved}{704}
\newcommand{\blocksComplexityWorsened}{63}
\newcommand{\blocksEntropyWorsened}{49}
\newcommand{\complexityEntropyWorsened}{21}
\newcommand{\blocksComplexityAvgEffect}{0.4744168293559271}
\newcommand{\blocksEntropyAvgEffect}{0.47360940741016105}
\newcommand{\complexityEntropyAvgEffect}{0.21551732345779007}
\newcommand{\dominatingSolutions}{74}
\newcommand{\threeObjectivesImproved}{74}
\newcommand{\threeObjectivesWorsened}{2}
\newcommand{\threeObjectivesAvgEffect}{0.38784785340795935}

\newcommand{\percentageProjectsImprovedComplexityEntropy}{\SI{70.4}{\percent}\xspace}

\newcommand{\meanBlocksBeforeRefactoring}{86.27427427427428}
\newcommand{\meanBlocksAfterRefactoring}{84.53226559893226}

\newcommand{\meanBlocksBeforeRefactoringMean}{86.27427427427428}
\newcommand{\meanBlocksAfterRefactoringMean}{89.2686217301114}
\newcommand{\blocksAfterRefactoringP}{< 0.001}

\newcommand{\meanComplexityBeforeRefactoring}{3.3636309183391457}
\newcommand{\meanComplexityAfterRefactoring}{2.08291137174689}
\newcommand{\complexityAfterRefactoringP}{< 0.001}

\newcommand{\meanEntropyBeforeRefactoring}{1.4447769417171383}
\newcommand{\meanEntropyAfterRefactoring}{1.1932980127859478}
\newcommand{\entropyAfterRefactoringP}{< 0.001}

\newcommand{\populationSize}{30}
\newcommand{\numGenerations}{100}
\newcommand{\maxRunTime}{1800}

\newcommand{\meanfrontsize}{21.466295506067475}

\newcommand{\longscriptsDiff}{165.16666666666674}
\newcommand{\longscriptsP}{< 0.001}
\newcommand{\scriptsP}{< 0.001}
\newcommand{\projectsWithDominatingSolution}{354}

\newboolean{inputappendix}
\setboolean{inputappendix}{true}

\begin{document}

\title{Improving Readability of Scratch Programs with Search-based Refactoring}

%\author{%
%  \IEEEauthorblockN{Anonymous Author(s)}%
%}

\author{%
\IEEEauthorblockN{Felix Adler, Gordon Fraser, Eva Gründinger, Nina Körber,\\ Simon Labrenz, Jonas Lerchenberger, Stephan Lukasczyk, Sebastian Schweikl}%
  \IEEEauthorblockA{%
    University of Passau\\%
    Passau, Germany%
  }%
}

\maketitle              % typeset the header of the contribution
\begin{abstract}
  %
  % Context
  %
  Block-based programming languages like \Scratch have become
increasingly popular as introductory languages for novices.
  % Problem
  %
These languages are intended to be used with a ``tinkering'' approach which allows learners and teachers to quickly assemble working programs and games, but this often leads to low code quality. Such code can be hard to comprehend, changing it is error-prone, and learners may struggle and lose interest.
The general solution to improve code quality is to refactor the code. However, \Scratch lacks many of the common abstraction mechanisms used when refactoring programs written in higher programming languages.
%
%
  % Insight/contribution
  %
In order to improve \Scratch code, we therefore propose a set of atomic code transformations to optimise \emph{readability}
by (1)~rewriting control structures and (2)~simplifying scripts using the inherently concurrent nature of \Scratch programs.
By automating these transformations it is possible to explore the space of possible variations of \Scratch programs.
In this paper, we describe a multi-objective search-based approach that determines sequences of code transformations
which improve the readability of a given \Scratch program and therefore form refactorings.
  % Results
  %
Evaluation on a random sample of \num{1000} \Scratch programs demonstrates that
the generated refactorings reduce complexity and entropy in
\percentageProjectsImprovedComplexityEntropy of the cases, and 
%often (\dominatingSolutions~projects) also reduce size at the same time. 
\projectsWithDominatingSolution~projects are improved in at least one metric without making any other metric worse.
The
refactored programs can help both novices and their teachers to improve their
code.
\end{abstract}

\begin{IEEEkeywords}
  Search-Based Refactoring, Scratch, Readability
\end{IEEEkeywords}

\section{Introduction}\label{sec:introduction} % Jonas

\begin{figure}[t]
    \centering
    \begin{subfigure}[t]{0.49\linewidth}
        \centering
        \includegraphics[scale=0.16]{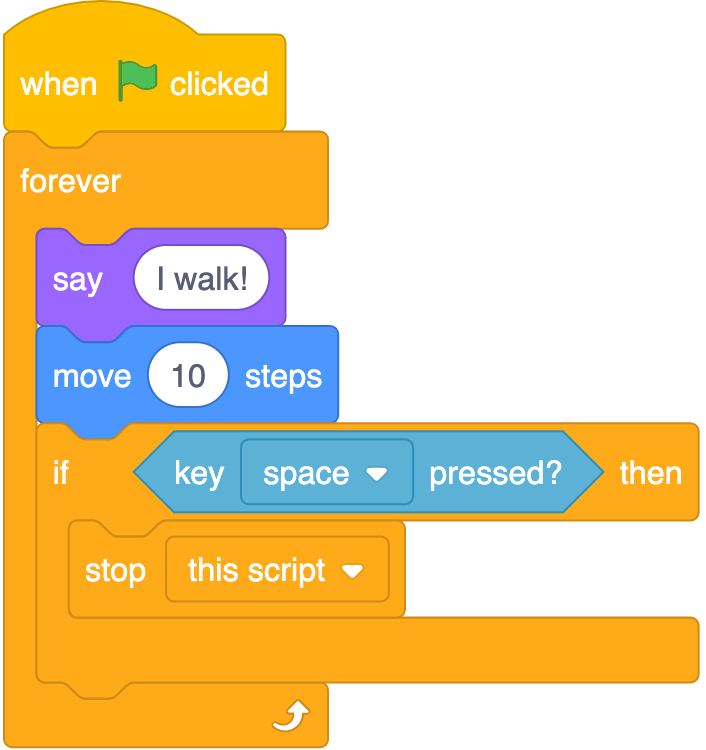}
        \caption{Unnecessarily complex code}
        \label{fig:example_bad}
    \end{subfigure}
    \hfill
    \begin{subfigure}[t]{0.49\linewidth}
        \centering
        \includegraphics[scale=0.16]{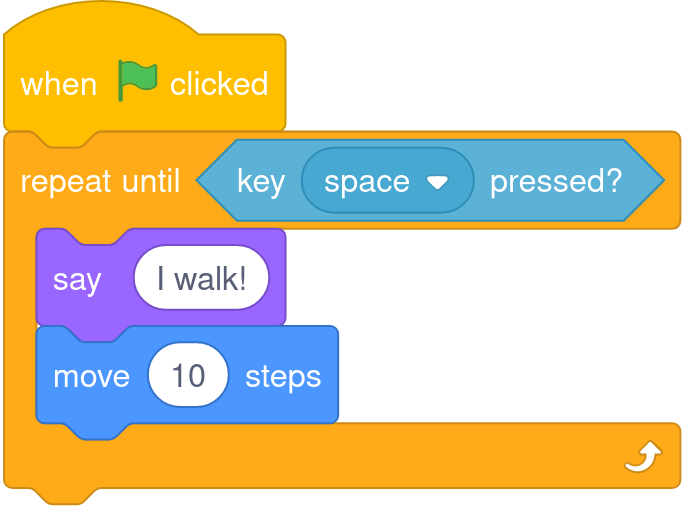}
        \caption{Refactored version}
        \label{fig:example_good}
    \end{subfigure}
    \caption{\label{fig:intro_example}The same functionality can be written in different ways, resulting in code that is easier or more difficult to read.}    
    \vspace{-1em}
\end{figure}

\Scratch~\cite{maloney2010} is a block-based programming language created
to introduce novices to the world of programming in a fun way.
The shapes of the blocks ensure that only syntactically valid code can be assembled, and high-level programming statements make it easy and quick to create working programs and games.
Programming with \Scratch is usually learned in a self-directed way~\cite{maloney2010} or taught by instructors who often are not skilled programmers themselves. As a result, \Scratch programs tend to have low code quality~\cite{techapalokul2017understanding},
which in turn has been found to negatively impact the pedagogical
effectiveness~\cite{HA16, techapalokul2017understanding}.
For example, \cref{fig:example_bad} contains a \Scratch script in which a sprite is controlled in a loop. While functionally correct, the use of a loop-condition nested in an if-block makes the code unnecessarily complicated.

In software engineering, code with low quality is typically \emph{refactored}~\cite{Fowler1999}, which means that the design is improved without changing the implemented behaviour, for example by extracting or moving methods. Such common refactorings tend to rely on abstraction mechanisms that are only available in higher programming languages~\cite{TT19}, but not in \Scratch. Even when they are available, their use might not be desirable in the context of young programming learners who are already busy trying to understand the most basic programming concepts. 
However, it is still possible to improve \Scratch code using refactorings tailored for the specifics of the \Scratch programming language. In particular, in this paper we focus on refactorings intended to improve the \emph{readability} of the code, which is directly linked to its understandability~\cite{Posnett2011}.
For example, \cref{fig:example_good}, shows a refactored and more readable version of \cref{fig:example_bad} which has the same functionality, yet uses fewer blocks, less complex control flow, and overall just looks tidier.

At this level of granularity, even small learners' programs may offer overwhelmingly many opportunities to apply such refactorings.
In order to support programming learners and their teachers, we propose an automated approach to identify sequences of changes to programs
that lead to an overall improvement in readability and therefore form refactorings.
We define a set of atomic \Scratch code transformations based on rewriting control structures as well as the event-driven distribution of
code to concurrent scripts. All implemented refactorings are designed to preserve program semantics
by respecting important dependencies such as data, time and control~\cite{Gupta2015}.  Given a candidate program, we then use a
meta-heuristic search algorithm to navigate the search space of possible transformation sequences in order to find versions of
the program that reduce its complexity, entropy and size, which are three important factors that have been established to influence code
readability~\cite{Posnett2011}.

In detail, the contributions of this paper are as follows:
\begin{itemize} %TODO: related work?
%    \item We give the technical background (Section~\ref{sec:background}).
    \item We propose a set of \numtransformations\ atomic code transformations for \Scratch programs (\cref{sec:transformations}).
	\item We introduce a search-based approach to generate sequences of transformations that improve the readability of \Scratch programs  (\cref{sec:fitness-functions,sec:representation,sec:search-operators,sec:algorithm}).
  \item We evaluate an implementation of this approach on a random sample of \num{1000} learners' programs (\cref{sec:evaluation}).
\end{itemize}

Our experiments demonstrate that search-based refactoring improves \Scratch
programs: Out of \num{1000} projects, \complexityEntropyImproved~are improved
with respect to complexity and entropy, and
\projectsWithDominatingSolution~projects result in dominating solutions, i.e.,
programs that are better in at least one metric, and not worse in any.
%
%. While size is more challenging to
%improve, it is not a primary target of our transformations.
%
Our approach is implemented as part of the \litterbox~\cite{litterbox} analysis
framework for \Scratch programs, and is freely available to support learners,
teachers, and researchers.

\section{Background}\label{sec:background}
\subsection{Search-based Refactoring} % Eva

Refactoring describes the process of improving code quality without changing
functionality. Concrete refactorings, i.e., generic and re-usable steps to
alter code with the intent to improve its quality, are often defined in order
to remove code smells~\cite{Fowler1999}. For example, duplicated code
can be refactored by replacing repeated segments of code with calls to an
extracted function capturing the common functionality. Many common refactorings
can be automated, and software developers have a wealth of different automated
refactorings at their disposal in modern integrated development
environments~\cite{KZN12,RPB12}. However, developers still face the challenge of
having to decide when and where to apply which refactoring.
Search-based refactoring~\cite{OC03} aims to address this challenge by
exploring the search space of possible program refactorings for a given
program, guided by fitness functions that measure aspects of code or design
quality. Since it is difficult to capture code quality with a single metric, it
is common to use multiple different metrics and multi-objective search
algorithms when applying search-based refactoring~\cite{HT07}. It has been
shown that this approach can successfully lead to improvements~\cite{OC08,OKS+16,AKM+20}, and the field is an active area of
research~\cite{MT04,MG18,MARIANI201714}.

\subsection{Code Quality Analysis for \Scratch}

%
%\Scratch~\cite{maloney2010} is a block-based programming language commonly used
%to introduce novices to the concepts of programming.
In \Scratch, programs are created by dragging and dropping puzzle-like \emph{blocks} in the \Scratch editor\footnote{\url{https://scratch.mit.edu/projects/editor/}, last accessed 2021–06–02.}.
In total, there are over one hundred blocks to choose from\footnote{\url{https://en.scratch-wiki.info/wiki/Blocks}, last accessed 2021–06–02.}.
Blocks connected to each other form \emph{scripts}. Every script
belongs to either the background of the game, the so called \emph{stage},
or to a \emph{sprite}, i.e., an object acting on the stage.
The first block of a script usually is a \emph{hat block}, i.e., an event listener that triggers execution of the script.
Blocks have different visual shapes, allowing only grammatically valid combinations of blocks.
In the case of hat blocks, for example, blocks can only be added at the bottom,
but not at the top. Control blocks like if-statements or loops can enclose
other blocks, and blocks terminating control flow prevent users from
adding subsequent blocks. % TODO do we need to go into further detail here? Expressions? Custom Blocks?

% More detailed formalizations of \Scratch have been proposed~\cite{whisker,VerifiedFromScratch}. % TODO maybe add anomaly detection? We have some kind of formalization there, as well...?
Although \Scratch is built to prevent syntax errors, programmers can still
struggle to write code of reasonable quality. They can develop
negative coding habits~\cite{meerbaum2011habits} and introduce code smells to
their code~\cite{aivaloglou2016kids,hermans2016a,techapaloku2017b, robles2017software}.
These smells make it harder to understand the code~\cite{HA16} and
might lead to bugs when the code is edited later on.
Several static analysis tools have been proposed to find code smells in \Scratch programs, for example \textsc{Hairball}~\cite{boe2013hairball}, \textsc{Quality
hound}~\cite{techapaloku2017b} or
\textsc{SAT}~\cite{chang2018scratch}.
\litterbox~\cite{litterbox} also detects code smells as well as instances
of general, predefined bug patterns automatically~\cite{bugpatterns}.

Since code smells are common in \Scratch, an obvious solution would be to
transfer the idea of automated refactoring to \Scratch. However, this is
challenging because many of the abstraction mechanisms underlying common
refactorings (e.g., inheritance, classes, methods) are not available in
\Scratch, or only in very basic form.
Although some common extraction refactorings can be approximated by extracting
``custom blocks'' and by simulating inheritance through the concept of
``clones'' in \Scratch~\cite{TT19}, these are advanced concepts that may
overwhelm early learners.
Transferring the idea of search-based refactoring to \Scratch faces further
challenges, as search-based refactoring techniques traditionally use
object-oriented design metrics (e.g., cohesion, coupling) to guide the search.
Since \Scratch is not an object-oriented language, other metrics are
necessary to guide search-based refactoring.

% Another approach to assess the quality of \Scratch programs is to run automated
% tests. The \toolname{Itch} testing framework does that by first translating the
% program under test to Python code and then running tests against the Python code.
% Its application, however, is limited to \scratchblock{say} and \scratchblock{ask}
% blocks. In contrast to that, \whisker~\cite{whisker} is a testing framework which
% executes the tests directly in the \Scratch IDE.
% A more general program analysis is provided by \bastet~\cite{VerifiedFromScratch},
% which enables for example software model checking on \Scratch programs.

\subsection{Code Readability}\label{sec:readability}

Considering the context of programming education, an alternative perspective on
code quality is how the code affects program comprehension. While difficult to
quantify, \emph{readability} of source code intuitively describes how easy it
is to understand it. %As defining an exact notion of readability is challenging,
Buse and Weimer~\cite{Buse2010} created a model of code readability based on
subjective human judgements of given code snippets, and demonstrated that this
metric strongly correlates with different aspects of code quality. The model is
based on a collection of syntactic features such as line length or types of
tokens used.
Posnett et al.~\cite{Posnett2011} demonstrated that this readability model can
be explained in terms of only three essential features: size, complexity, and
entropy. They used the common metric of lines of code to measure size, the
Halstead metric suite to quantify complexity, and entropy at the level of
syntactic tokens. The resulting model outperformed the original model of Buse
and Weimer, and the relevance of these properties was also confirmed
independently by Choi et al.~\cite{choi2018measuring}.
While there are many other attributes of code that affect comprehension, such
as variable names or other textual features~\cite{scalabrino2016improving}, a
viable initial step towards automatically improving readability therefore lies
in considering and optimising some metrics representative of size, complexity,
and entropy.

\section{Approach}\label{sec:approach}
\subsection{Code Transformations for \Scratch}\label{sec:transformations}

%\subsubsection{Modifying an AST with a list of refactorings}
\label{sec:ast}

In order to refactor \Scratch projects to more readable versions, we aim to
find sequences of code transformations which, when applied together, improve
the program. To this end, we define atomic transformations on the 
abstract syntax tree (AST) of \Scratch programs. Each transformation can be 
applied to individual nodes, subtrees, or the edges between these. 
A transformation takes the AST of a program~$S_n$ as input,
transforms it accordingly, and returns the modified AST representing the new
program version $S_{n+1}$ as output. 
For every transformation we also maintain information about where it is
applied to. An atomic transformation can potentially be applied at different
locations of the AST depending on its structure, but transformations may
require certain preconditions to hold in order to be applicable. In order to
determine for a given program $S_n$ which concrete transformations are
possible, we define the function \( \mathsf{findPossibleTransformations}(S_n) \)
for each type of transformation, which operationalises the matching of suitable
locations in the AST as well as the preconditions of the transformation and
returns a list of all possible instantiated transformations, applicable to the
AST $S_n$.

For most code transformations we also define their inverse transformations
($\rightleftarrows$) to enable the search to reach relevant intermediate states
of program transformations. In particular, we expect that these transformations
enable the search to escape local optima and potentially enable more powerful
transformations in the subsequent search.
In total, we define \num[round-precision=2]{\numtransformations} atomic transformations which are categorised as
either (1)~control flow transformations or (2)~concurrency transformations:

\emph{Control flow transformations}
%,
%listed in~\cref{subsec:control-transformations},
transform an individual script by reordering its blocks or replacing control
blocks by equivalent combinations of blocks. Control blocks can be nested and
hard to read, but can often be simplified as the abstraction level of the
\Scratch blocks varies, e.g., a \scratchblock{forever} loop which contains a
conditional termination can be simplified to a \scratchblock{repeat until}
loop. Furthermore, for conditionals, we apply transformations based on logical
equivalences of their conditions. We define the following control flow
transformations:
%
%We list the names/names of the inverses and a short descriptions of all transformations we defined here. Descriptions do not include descriptions of the inverse.
%\subsection{Atomic control flow transformations for \Scratch programs}\label{subsec:control-transformations}
{
\setlength{\parindent}{0em}
\begin{asparadesc}
    \item[Swap Statements] Swap two statements that are independent of each other, if swapping does not create new dependencies.
    \item[Loop Unrolling]  ($\rightleftarrows$ Sequence to Loop) Unroll a \scratchblock{repeat times} loop by repeating its body.
    \item[Forever If to Forever Wait]($\rightleftarrows$ Forever Wait to Forever If) Replace an \scratchblock{if} block inside a \scratchblock{forever} loop with a \scratchblock{wait until} block with the same condition.
    \item[Extract Loop Condition] ($\rightleftarrows$ Inline Loop Condition) Transform a \scratchblock{forever} loop that conditionally terminates the script or the program to a \scratchblock{repeat until} loop.
    \item[Split If Body] ($\rightleftarrows$ Merge Double If) Split the body of an \scratchblock{if} block. Replace the \scratchblock{if} block by one containing the first part of the body, add another \scratchblock{if} for the remaining statements.  %of the body.
    \item[If Else to If If Not] ($\rightleftarrows$ If If Not to If Else) Split an \scratchblock{if else} block into two \scratchblock{if} blocks. The second \scratchblock{if} block checks on the negated initial condition.
    \item[Ifs to Conjunction]  ($\rightleftarrows$ Conjunction to Ifs) Transform two nested \scratchblock{if} blocks into an \scratchblock{if} block which checks for the conjunction of the initial conditions.
    \item[If If Else to Conjunction]  ($\rightleftarrows$ Conjunction to If If Else) Replace an \scratchblock{if} containing an \scratchblock{if else} by two \scratchblock{if} blocks. The condition of the first \scratchblock{if} is the conjunction of the two initial conditions; the second condition is the one of the first if.
    \item[If Else to Disjunction] ($\rightleftarrows$ Disjunction to If Else) Replace an \scratchblock{if} block in the else case of an \scratchblock{if else} block by an \scratchblock{if} with the disjunction of the two conditions if the then cases of the initial conditionals have the same statements.
\end{asparadesc}%
}

\emph{Concurrency transformations}
%,
%listed in~\cref{subsec:concurrency-transformations},
are based on the event-driven nature of \Scratch programs. For example, it is common practice to
place independent functionality in separate concurrent scripts. Consequently, sometimes it is possible to split loops and scripts
into several smaller scripts, which are executed concurrently. For instance, the
\emph{Extract Independent Subscripts} transformation splits a script into
multiple independent scripts. In order to preserve the semantics of the program
it is important that dependencies between statements are considered when
deciding which transformations can be applied:
\begin{itemize}
	\item \emph{Control dependencies:} We consider control dependencies using a classical control dependence graph; it is not possible to split statements if one is control dependent on the other. 
	\item \emph{Data dependencies:} We consider data dependencies by building a data dependence graph based on a classic reaching definitions analysis. As an adaptation to \Scratch, this analysis has to take not only the variables in the program into account, but also the attributes of the sprites and the stage. In particular, for each sprite we consider its position, rotation, costume, size, and visibility as attributes, and define for each of the program statements in \Scratch whether it defines or uses this attribute. 
	\item \emph{Time dependencies:} \Scratch programs tend to make heavy use of timing-related statements, for example to control the speed of movement of sprites, to synchronise interactions between sprites, or to encode the steps of sequences of animations or interactions. If a statement is a successor of a timing-related statement, then it is not possible to split the script between these statements as the concurrent execution would not adhere to the same timing. For each block in the \Scratch language we determined whether it is timing-related, and we use a simple forward-may dataflow analysis to identify which statements are time-dependent on which other statements.
\end{itemize}

%The result of the search is a sequence of code transformations. To guide
%the search towards sequences which form valuable refactorings, we need \emph{fitness functions}
%to evaluate the quality of a potential refactoring.
We define the following concurrency transformations:
%
%\subsection{Atomic transformations based on the inherently concurrent nature of \Scratch programs}\label{subsec:concurrency-transformations}
{
\setlength{\parindent}{0em}
\begin{asparadesc}
    \item[Split Loop] ($\rightleftarrows$ Merge Loops) Split the body of a loop if there are no dependencies between the statements of its body and splitting does not create new ones. Replace the initial loop body by the first part of the body, add another loop for the remaining statements of the body.
    \item[Split Script]($\rightleftarrows$ Merge Scripts) Split a script if there are no dependencies upwards.
    \item[Extract Independent Subscripts] Split a script with dependencies into new scripts
    which do not depend on each other but respect the dependencies in the initial script. We define no inverse transformation since it would produce too many options of which scripts to combine and in which order.

    \item[Extract Events from Forever] ($\rightleftarrows$ Merge Events into Forever) Replace a \scratchblock{forever} loop with \scratchblock{key pressed}
        conditionals by scripts triggered by \scratchblock{when key pressed} event listeners.
    \item[Split Script after Until] ($\rightleftarrows$ Merge Scripts after Until) Split a script after a
        \scratchblock{repeat until} loop. Add a new script with a \scratchblock{wait until}
       and the same condition.
\end{asparadesc}%
}

\subsection{Fitness Functions}\label{sec:fitness-functions}

The aim of the code transformations is to improve the readability of the code.
In order to guide the search to achieve this objective, we require fitness
functions that encode relevant aspects of code quality. A common approach in search-based refactoring is to use different metrics in a multi-objective optimisation scenario.
Intuitively, we would like to improve the readability of \Scratch programs by
avoiding unnecessary complexity, keeping programs as small as possible, and
optimising coherence of the code within individual scripts. These objectives
are reminiscent of work on modelling subjective code readability, which has
been shown to be influenced by size, complexity, and
entropy~\cite{Posnett2011}. Consequently, we define the following fitness
functions:

\subsubsection{Total number of blocks}\label{sec:number_of_blocks}
In order to keep solutions as simple as possible and to avoid that the search
unnecessarily inflates programs, one goal of optimisation is to minimise the
size of programs. We measure the size of a \Scratch program in terms of the
blocks it consists of. Blocks can represent not only statements but also
expressions. For example, the script in \cref{fig:example_bad} consists of nine
blocks: An event block, the forever loop, the say and move statements, the if
statement, the stop statement, the sensing block checking if the space key has
been pressed, and the two drop-down boxes are also counted as blocks since they
can be replaced with other blocks. The refactored script in
\cref{fig:example_good} only consists of six blocks.

\subsubsection{Block Category Entropy}\label{sec:entropy}
The concept of entropy is used in information theory to describe the
uncertainty or surprise in a random variable, and can be thought of as the
amount of information contained in the variable. A higher entropy describes a higher uncertainty of a variable. It has been shown that the
entropy of the tokens in source code is directly related to the readability of the code~\cite{Posnett2011}, and we therefore aim to minimise it. We measure entropy at the level of blocks in a script. The blocks in the \Scratch programming language are organised in different categories depending on which aspects of functionality they address. The main categories are motion, looks, control, sensing, operators, variables, and events. We calculate entropy in terms of the categories of blocks within a script, which intuitively means that a script has low entropy if it is only responsible for one type of functionality. The category entropy for a script $S$ is calculated as Shannon entropy $H(S)$ given the number of blocks of category $c_{i}$ as $\text{\textit{count}}(c_{i})$, and total number of blocks as $\text{\textit{numberOfBlocks}}(S)$:
\begin{equation} \label{eq:entropy}
\begin{split}
H(S) & = - \sum_{i=1}^n p(c_{i}) \cdot \log_{2} p(c_{i}), \text{with}\\
p(c_{i}) & = \frac{\text{\textit{count}}(c_{i})}{\text{\textit{numberOfBlocks}}(S)}
\end{split}
\end{equation}
For example, the script in \cref{fig:example_bad} contains one event block, one looks block, one motion block, two sensing blocks, two control blocks, and the two menu blocks. As the total number of blocks is 9, the entropy is 
\begin{align*}
  H(S) = -\big(
             &\tfrac{1}{9} \cdot \log_{2} \tfrac{1}{9} + 
              \tfrac{1}{9} \cdot \log_{2} \tfrac{1}{9} + 
              \tfrac{1}{9} \cdot \log_{2} \tfrac{1}{9} + {} \\
             &\tfrac{2}{9} \cdot \log_{2} \tfrac{2}{9} + 
              \tfrac{2}{9} \cdot \log_{2} \tfrac{2}{9} +
              \tfrac{2}{9} \cdot \log_{2} \tfrac{2}{9}
           \big)
       = 2.50.
\end{align*}
In contrast, the refactored script in \cref{fig:example_good} contains only one block of each type, such that 
\begin{align*}
  H(S) = -\big( 
            &\tfrac{1}{6} \cdot \log_{2} \tfrac{1}{6} + 
             \tfrac{1}{6} \cdot \log_{2} \tfrac{1}{6} + 
             \tfrac{1}{6} \cdot \log_{2} \tfrac{1}{6} + {} \\
            &\tfrac{1}{6} \cdot \log_{2} \tfrac{1}{6} + 
             \tfrac{1}{6} \cdot \log_{2} \tfrac{1}{6} +
             \tfrac{1}{6} \cdot \log_{2} \tfrac{1}{6}
          \big)
       = 2.58.
\end{align*}
Thus, the refactored script has a higher entropy as there is higher uncertainty about the categories of blocks used.

Since we aim to improve the constituent scripts of a program, as fitness function we compute the average category entropy for each script in the
program.

\subsubsection{Complexity}\label{sec:complexity} 

The Halstead suite of metrics intends to quantify different complexity-related properties of a program such as volume, difficulty, or effort. The metrics are calculated using information about the operators and operands used in the program. As operators we count all blocks representing statements, events, and blocks from the ``operators'' category, while we define literals, variables, parameters, and drop-down menu options as operands. We use the Halstead difficulty as target metric to optimise, as it intends to quantify how easy it is to understand a program while reading or programming. It is calculated as follows:
\begin{equation} \label{eq:difficulty}
D = \frac{\# \text{unique operators}}{2} \cdot \frac{\# \text{total operands}}{\# \text{unique operands}}
\end{equation}
%
%
% The program volume $V$ is the information contents of the program.
% In line with previous work~\cite{Posnett2011} we use the Halstead Volume in our approach to optimise.
% The calculation of the volume requires a definition of operations and operands.  The different properties of the scratch blocks are divided.  For example, statements and expressions are assigned to operations,  literals and attributes to operands.
% Further now we compute the unique operands and operators, as well as the total number of operands and operators. This provides us with the length and size.
% The volume is now calculated as follows.
% \begin{equation} \label{eq:volume}
% \begin{split}
% length & = \#totalOperands + \#totalOperators \\
% size & = \#UniqueOperands + \#UniqueOperators \\
% V &  = length \cdot \log_{2}{size}
% \end{split}
% \end{equation}
%
For example, the script in \cref{fig:example_bad} consists of seven operators (the blocks) and four operands (the two literals and the two drop-down menu options), all of which are unique. Consequently, the Halstead difficulty is $\frac{7}{2} \cdot \frac{4}{4} = 3.5$. The refactored script in \cref{fig:example_good} consists of five operators and three operands, and thus has a lower Halstead difficulty of $\frac{5}{2} \cdot \frac{3}{3} = 2.5$.
Similar to \cref{sec:entropy} we compute the average Halstead difficulty per script and use this as fitness value for the optimisation.

\subsection{Refactoring Representation}\label{sec:representation} % Simon

\label{sec:grammatical} % Simon

In order to enable the search to find program transformations, a
suitable representation is required for these
transformations. One possibility would be to apply the search directly on
syntax trees in the style of genetic programming, and thus to generate a
completely transformed AST as result of the search. However, presenting 
a modified program that may be very different from its original version to the user without
explanation of how this was derived is not acceptable
for our use case. Consequently, we need the search to evolve sequences of
refactorings that explain the changes. However, storing concrete code
transformations as lists would create the following problem:
Each transformation in a sequence depends on the state of the program after the previously executed alteration. Standard search operators such as mutation and crossover (\cref{sec:search-operators})
may thus break individuals of the search if the state of the AST has changed as part of other transformations (for example, a merge transformation in a sequence may no longer be applicable if a prior transformation removes one of the merged scripts).

To overcome this issue, we use an integer representation inspired by
grammatical evolution~\cite{o2001grammatical}.
In grammatical evolution, the genotype is given by a list of integers
(\emph{codons}). The phenotype is obtained by applying a mapping as follows:
Starting at the first production of starting symbol $n_s$ for a given grammar,
we choose the $r$th production rule out of $n$ available rules for a current
non-terminal $x$. For a single codon $c$ the chosen production rule $r$ is
calculated as follows:
\begin{equation*}
  r  = c \; mod \; n
\end{equation*}
When a production is selected, the next codon is decoded. If no more codons are
left, or for one state of program the set of production rules is empty, the
mapping stops.

Conceptually, the grammar to produce a sequence of transformations of length
$n$ is given by the following grammar:
\[
  \begin{aligned}
    \text{\textit{solution}}  ::= \; & \text{\textit{transformation}}_1 \; \text{\textit{transformation}}_2 \; \\
    \ldots \;        & \text{\textit{transformation}}_n
  \end{aligned}
\]
The terminals of our grammar are individual transformations of the AST
applicable to the given state of a program. 
%Each transformation stores unique
%nodes, edges or subtrees of the AST that are modified by the transformation.
%
The following production defines the possible transformations for a given state
of a program~\( S_{n} \):
\[
  \begin{aligned}
    \text{\textit{productions}} ::= \; & \text{\textsf{findPossibleTransformations}}(S_n)
  \end{aligned}
\]
The resulting program state $S_{n+1}$ is returned by applying
the transformation.
With this representation we do not need to store lists of concrete
transformations, but the genotype is a simple list of integers that encode the
applied production rules. Consequently individuals always represent valid
sequences of transformations, and it is straightforward to define search
operators.
Note, however, that the same integer %in a list of chosen production rules in one solution
may represent different AST transformations in different solutions.

\begin{figure*}[t]
  \begin{subfigure}[t]{.5\textwidth}
    \centering
    \includegraphics[scale=0.16]{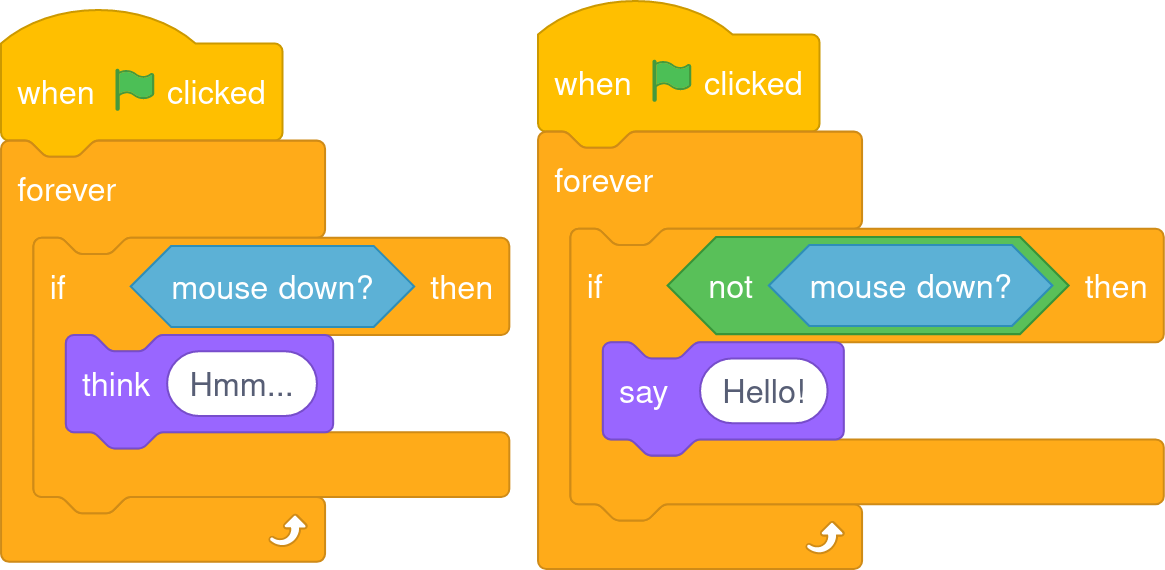}
    \caption{Original Program~\( S_{0} \)}
    \label{fig:two-possible-refactorings}
  \end{subfigure}
  \begin{subfigure}[t]{.25\textwidth}
    \centering
    \includegraphics[scale=0.16]{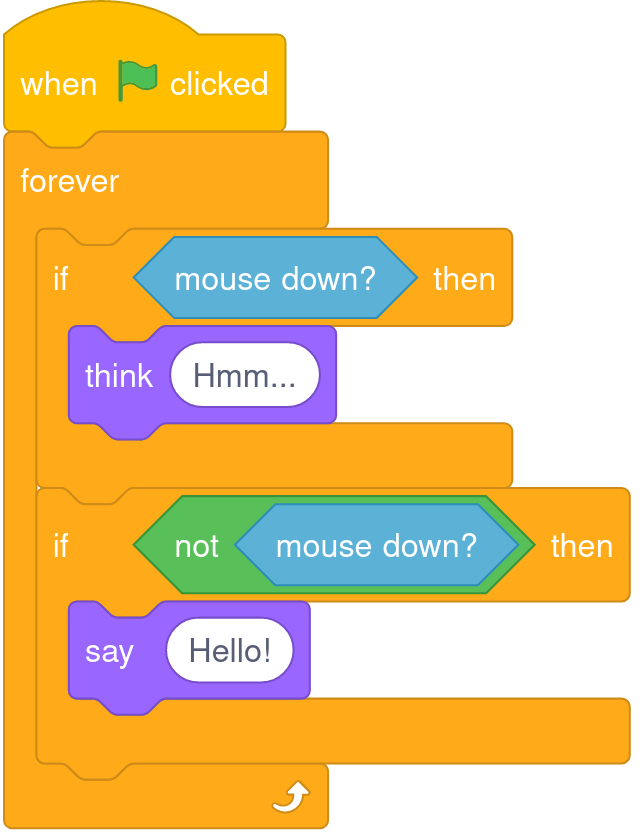}
    \caption{Program~\( S_{1} \)}
    \label{fig:merged-loops}
  \end{subfigure}
  \begin{subfigure}[t]{.24\textwidth}
    \centering
    \includegraphics[scale=0.16]{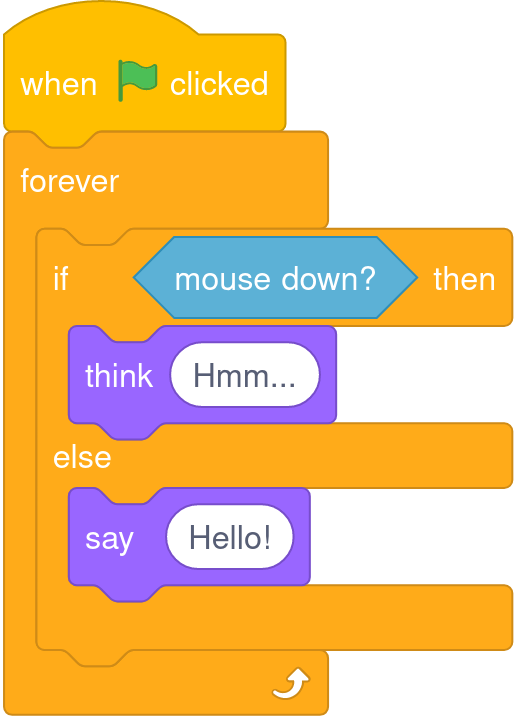}
    \caption{Final program~\( S_{2} \)}
    \label{fig:final-program}
  \end{subfigure}
  \caption{%
    Example program transformation.
    Starting with the initial program variant \( S_{0} \),
    we obtain \( S_{1} \) by merging the two forever
    blocks in \( S_{0} \). 
    Then, the two if blocks in \( S_{1} \) are merged
    to arrive at the final program variant \( S_{2} \).
  }
  \label{fig:transformation-example}
\end{figure*}

\paragraph*{Example} Consider the program $S_0$ in
\cref{fig:two-possible-refactorings}, and suppose
$\textsf{findPossibleTransformations}(S_0)$ returns the following possibilities:
\begin{compactenum}
  \item replace the left \scratchblock{forever} \scratchblock{if} blocks with a \scratchblock{wait until} block
  \item replace the right \scratchblock{forever} \scratchblock{if} blocks with a \scratchblock{wait until} block
  \item merge the two loops
\end{compactenum}
Now, let the following list of integers be given as an example encoding of a
phenotype for the program~\( S_{0} \):
\[
  \begin{aligned}
    T  = \; & \langle  \; 5 \; 9 \; 10 \; 42 \; 17\; 8 \; 13 \; 2 \; 13 \; \rangle
  \end{aligned}
\]
The decoding would start with the codon~$5$.
Since the described three productions are applicable for the initial program,
the decoding to the first transformation thus looks as follows:
\[
  \begin{array}{lcl}
    5 \; mod \; 3 = 2 & \;\longrightarrow\;
                      & S_1 = \textsf{merge\_loops($S_0$) } \\
  \end{array}
\]
Since the calculated production rule was $2$, the third option of all possible
productions for the given state is chosen ($0$ would have been the first option).
Afterwards the newly created program $S_1$ (seen in \cref{fig:merged-loops}) is
evaluated again by computing \( \textsf{findPossibleTransformations}(S_1) \).
Suppose this results in just one possible production, the \emph{If If Not to If
Else} transformation of the two consecutive \scratchblock{if} blocks. So the
next decoding of our codon with the number 5 would look as follows:
\[
  \begin{array}{lcl}
    9 \; mod \; 1 = 0 & \;\longrightarrow\;
                      & S_2 = \textsf{ififnot\_to\_ifelse($S_1$)} \\
  \end{array}
\]
This creates the new state $S_2$, seen in \cref{fig:final-program}.
Assuming $\textsf{findPossibleTransformations}(S_2)=\emptyset$ the evaluation
of our solution $T$ is finished, with $S_2$ being the final state of the
refactored program that can be evaluated by its fitness.

\subsection{Search operators}\label{sec:search-operators} % Simon

%\subsubsection{Length of a solution}\label{sec:length}
The number of required code transformations is unknown a priori
and can be different for each program. We therefore chose a variable length for 
the encoding and rely on the search to find a suitable length for each solution.
%
%\subsubsection*{Generating a solution}\label{sec:generator}
For each randomly generated individual of our initial population, we first
select a random number $n$ in the range $[1..max]$, with $max$ being the
maximum number of codons in a phenotype. 
We then generate $n$ random codons by uniformly sampling integers in the
range of $0$ to an upper bound $x$.
As constraint, $x$ must be bigger than the maximum number of possible code
transformations, otherwise the decoding of a solution might never choose
productions with a higher number than $x$ due to the modulo operator.

%\subsubsection{Exploration}\label{sec:exploration}
% 
When mutating an individual of length $n$, each codon is modified with
probability $1/n$, by either (1)~replacing it with a random codon,
(2)~inserting a new codon at the location, or~(3)~deleting the codon.
We use single-point crossover, which would not have been directly applicable on
a list of code transformations, but is easy for integer lists.
%
%We chose these operators based on prior work on variable size
%search~\cite{fraser2012whole}.
\subsection{Algorithm}\label{sec:algorithm} %Simon

We evaluate the candidate program transformation sequences regarding three
fitness values with conflicting objectives (see \cref{sec:fitness-functions})
and therefore use the NSGA-II~\cite{deb2002fast} search
algorithm, which has been shown to be effective for many software engineering problems~\cite{harman2012search}.
%
% \begin{algorithm}\label{alg:nsga-ii}
%     \SetAlgoLined
%     $R_t$ = $P_t \cup Q_t$\;
%     $F_t$ = fast-non-dominated-sort($R_t$)\;
%     $P_{t+1}$ = $\emptyset$ and $i=1$\;
%     \While{$\left|P_{t+1}\right| + \left|F_i\right| > N$} {
%         crowding-distance-assignment($F_i$)\;
%         $i = i + 1$\;
%     }
%     Sort($F_i, \prec_n$)\;
%     $P_{t+1}$ = $P_{t+1} \cup F_i[1:(N-\left|P_{t+1}\right|)]$\;
%     $Q_{t+1}$ = make-new-pop($P_{t+1}$)\;
%
%     $t = t + 1$\;
%     \caption{NSGA-II Pseudo Code \cite{deb2002fast}}
% \end{algorithm}
%
%, even for conflicting
%objectives~\cite{deb2002fast}. 
The algorithm can optimise for conflicting objectives, due to its reliance on
Pareto dominance, which is defined as follows:
One solution $x_1$ dominates a second solution $x_2$ if $x_1$ is not worse in
any objective than $x_2$, and $x_1$ is strictly better than $x_2$ in at least
one objective~\cite{deb2002fast}. In other words, $x_1$ dominates $x_2$,
written as $x_1 \prec x_2$, when the following holds for objectives 
$f_{1}, \ldots, f_{M}$:
\begin{align*}
    \big( \forall & m \in \{1, \ldots, M\} : f_m(x_2) \leq f_m(x_1) \big)
      \enspace \land \\
    \big( \exists & m \in \{1, \ldots, M\} : f_m(x_2) < f_m(x_1) \big)
\end{align*}
All solutions are sorted into lists based on their Pareto dominance, the
so-called Pareto fronts. All solutions in the first front are not dominated by
another solution, all solutions in the second front are not dominated by any
other solution except for the ones in the first front, and so on.
The Pareto fronts help to determine which solutions are better than others, even
with conflicting objectives. They are specifically built in a way, that one
solution of the first front is definitely better than a solution in the second
front.
This means, with the NSGA-II and its \emph{fast-non-dominated-sort} of
populations, we have an algorithm that can separate lists of better and worse
solutions, even for conflicting objectives.
When deciding which solutions to include in the next population from
within a Pareto front, NSGA-II aims to improve diversity by sorting individuals according to the crowding
distance~\cite{deb2002fast}. 

NSGA-II runs until a stopping criterion is met. In our case, we stop the
evolution after a fixed number of generations or after a set threshold of seconds has passed.
The fixed value of generations sets a base for the comparison of solutions. The
timeout is important to cover for cases where particularly large programs or inefficient refactorings delay experiments.

%Summarized, we have three objectives, partially conflicting each other. Because
%of that, we have chosen the NSGA-II as our evolutionary algorithm, since the
%algorithm is well-suited for such a case~\cite{deb2002fast, harman2012search},
%where the number of objectives is not too high.

\section{Evaluation}\label{sec:evaluation}

In order to achieve a better understanding of the effectiveness of search-based
refactoring for \Scratch, we conducted an empirical analysis.
% to answer the following research questions:
%
A primary question is how often \Scratch programs can be improved using our
approach in the first place, therefore the first research question is as
follows:
%
% \begin{asparaitem}
% 	\item RQ1: How effective is search-based refactoring?
% 	\item RQ2: How do refactored \Scratch programs differ?
% 	\item RQ3: How are the \Scratch programs transformed?
% \end{asparaitem}

\begin{resq}[RQ1]\label{rq:effectiveness}
  How effective is search-based refactoring for \Scratch?
\end{resq}

The second research question aims to shed light on how the resulting programs
look like:

\begin{resq}[RQ2]\label{rq:readability-changes}
  How do refactored \Scratch programs differ?
\end{resq}

Finally, we would like to understand which transformations are used in order to derive the refactored programs:

\begin{resq}[RQ3]\label{rq:program-changes}
  How are the \Scratch programs transformed?
\end{resq}

\subsection{Experimental Setup}\label{sec:evaluation-setup}

An artifact that contains all data and software to reproduce our study is available online: \url{https://github.com/se2p/artifact-scam2021}.

\subsubsection{Implementation}\label{sec:evaluation-tool}

We used \litterbox~\cite{litterbox} to implement the search algorithm and transformations presented in this paper. \litterbox provides a parser that reads the
JSON-format representation of \Scratch programs and creates an AST.
Each transformation consists of two parts: % in the implementation: 
First, an AST-visitor encodes the matching and preconditions to derive
the concrete transformations. Second, a concrete transformation implements the
actual transformation for a specified target location. At the end of the
search, our prototype produces CSV statistics, and creates one modified
version of the \Scratch input file for each individual of the final Pareto
front.

\subsubsection{Experiment Subjects}\label{sec:evaluation-subjects}

We randomly sampled \num{1000} publicly shared \Scratch programs
 from the \Scratch website\footnote{%
  \url{https://scratch.mit.edu}, last accessed 2021–06–25.%
} between 2021–05–13 and 2021–06–10.
The projects were created between 2020–05–28 and 2021–03–13. We restricted
our sampling to programs with at least ten code blocks, to exclude projects
that are just art or contain no functionality. We furthermore excluded remixes,
which are modified and shared
versions of already uploaded \Scratch projects\footnote{%
  \url{https://en.scratch-wiki.info/wiki/Remix}, last accessed 2021–06–27.%
}, to ensure our dataset does not include the same code twice.

\subsubsection{Experiment Setting}\label{sec:evaluation-setting}

We executed \litterbox on each of the constituent programs in sequence
to apply the search-based refactorings.
For this,
we used a fork of \litterbox in Git revision \texttt{6b193f88}.
We conducted our experiments on dedicated computing machines,
each featuring two Intel Xeon E5-2620v4 CPUs
with \SI{2.1}{\giga\hertz} and \SI{256}{\giga\byte} of RAM.
The nodes are running Debian GNU/Linux~10.9 and OpenJDK~11.0.11.
We limit each execution of \litterbox
using the SLURM job scheduling system~\cite{YJG03}
to one CPU core and \SI[round-precision=1]{8}{\giga\byte} of RAM;
we set the available Java heap to \SI[round-precision=1]{6}{\giga\byte}.
We set the population size for NSGA-II to
\num[round-precision=2]{\populationSize}~chromosomes,
allowed a maximum number of \num{\numGenerations}~generations,
and a maximum run time for the search process of \SI{\maxRunTime}{\second}.

To answer RQ1, we compare the original projects with the Pareto front of refactored versions for each project and each run. We use the Vargha-Delaney $\hat{A}_{12}$ effect size to quantify the difference with respect to each of the metrics; when comparing improvement of multiple metrics we average the effect sizes of each of the constituent metrics. Since all objective functions are minimised, an effect size $< 0.5$ represents an improvement. We use a Wilcoxon rank sum test with $\alpha=0.05$ to determine when metrics are significantly improved.
To answer RQ2 we consider only the best individual within a Pareto front with respect to each metric. We use a Wilcoxon rank sum test to determine if differences are significant. 
To answer RQ3 we look at the distribution of code transformations contained in the solutions produced across all projects and runs.
%
%
%We give all numbers rounded to three significant digits,
%except for counted values where we report the exact count.
%

\subsection{Threats to Validity}\label{sec:evaluation-threats}

\subsubsection{Internal Validity}\label{sec:evaluation-threats-internal}

Meta-heuristic search is a randomised process, and different seed values for
the random-number generator can cause different results.
We therefore executed \litterbox \num[round-precision=2]{30} times on each
\Scratch project to mitigate the influence of randomness.
Although we carefully checked our implementation, bugs may always influence results.
Our transformations are semantics-preserving by design. To verify this,
we used \num[round-precision=2]{15} \Scratch projects from prior work for which
we have automated \whisker~\cite{whisker} tests. To verify that the code
transformations preserve the semantics, we executed these tests before and after
the search-based refactoring, ensuring that no tests change their outcome.

\subsubsection{External Validity}\label{sec:evaluation-threats-external}

We use \num{1000} \Scratch projects of different sizes
for our experiments.
The projects were randomly sampled
as described in \cref{sec:evaluation-subjects}.
As always with such sampling, our results might not generalise to other projects, and so replication studies will be important for future work.
%
%Furthermore,
%the \Scratch scheduler is inherently non-deterministic
%due to its parallel nature.
%
%This can also influence generalisability of our results.

\subsubsection{Construct Validity}\label{sec:evaluation-threats-construct}

We use NSGA-II~\cite{deb2002fast} as a search algorithm
and the total number of blocks,
block category entropy,
and complexity~(see \cref{sec:fitness-functions}) as search objectives.
These three metrics are generally accepted proxies that influence readability.
Other metrics may be better suited, and the approach can easily be adapted with other fitness functions.
We assume that splitting a script into smaller scripts improves readability,
which may not be the case for programming novices.
However, an evaluation of subjective readability will require a human study.
Although previous work showed NSGA-II to be effective~\cite{harman2012search},
different search algorithms or other parameter settings may influence the
achieved results.
%

%\subsection{Results}\label{sec:evaluation-results}

%We discuss the findings of our empirical evaluation in the following.

\subsection{RQ1: Effectiveness of Search-based Refactoring}\label{sec:evaluation-rq1}

\begin{figure}[t]
    \centering
  % Picture taken from project id: 400006216
    \begin{subfigure}[t]{0.59\linewidth}
        \centering
        \includegraphics[scale=0.155]{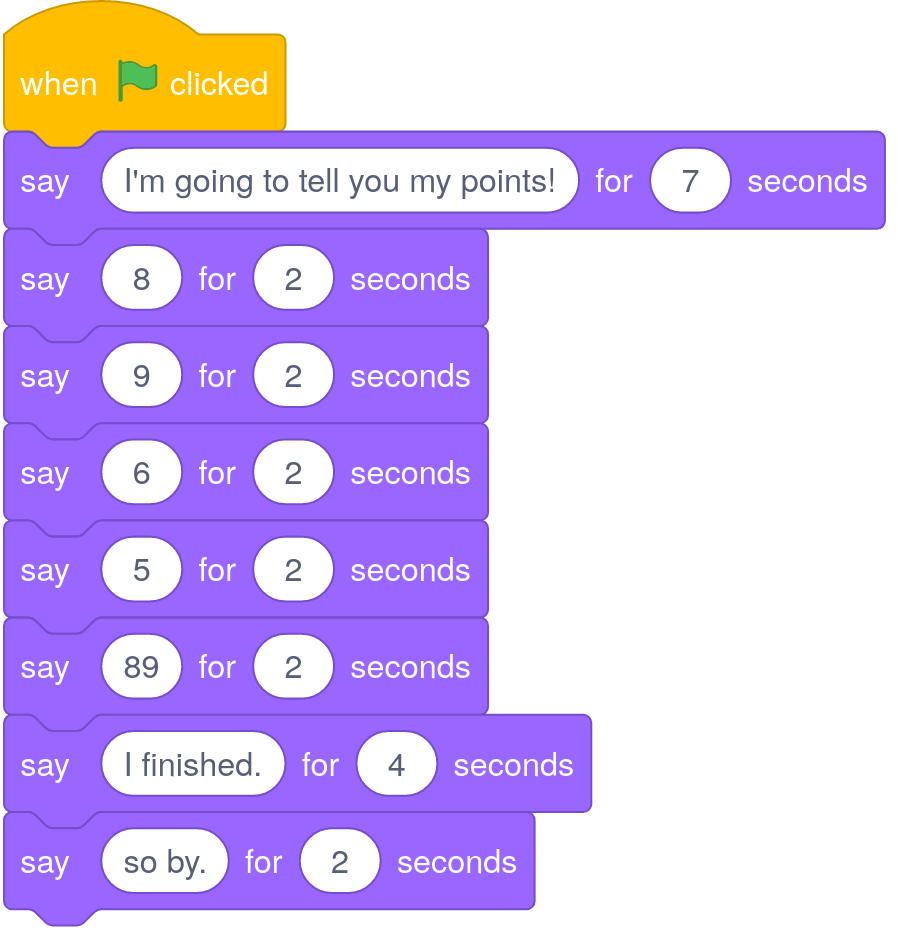}
        \caption{Timed sequences of statements}
        \label{fig:no-transformations-possible}
    \end{subfigure}
    \hfill
    %
    % Picture taken from project id: 402106862
    \begin{subfigure}[t]{0.39\linewidth}
        \centering
        \includegraphics[scale=0.155]{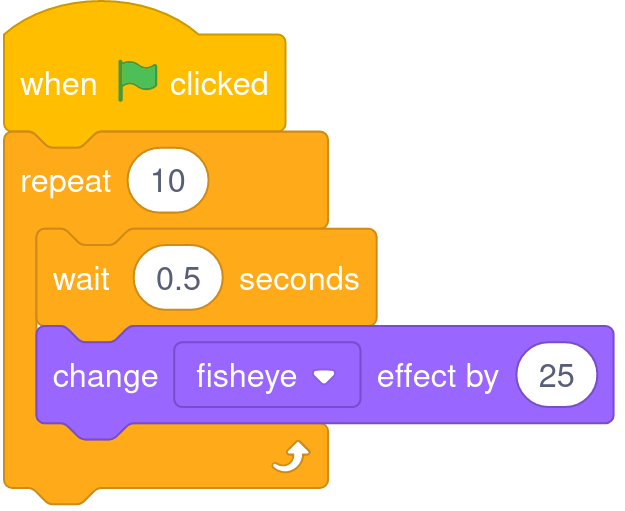}
        \caption{Animation with more complex control flow}
        \label{fig:no-transformations-possible-animation}
    \end{subfigure}
    \caption{Animations and story projects provide no opportunities for transformations.}
\end{figure}

Applicable code transformations were found for all but
\num[round-precision=2]{\projectsNotRefactored} projects in our dataset.
The projects which were not transformed are usually either too small, or simply
consist of only blocks with data- or time-dependencies to each other in
sequential order (e.g., \cref{fig:no-transformations-possible}).
In such a case, our transformations are not applicable as they would break the
original functionality of the program.
We conjecture that these projects are mostly animations or stories which
consist of sequences, no repetitions, and no opportunity for concurrency.
Indeed the number of scripts per project is noticeably lower for projects that
were not transformed (\num{\meanScriptCountNonRefactorable} on average)
compared to those that were (\num{\meanScriptCountRefactorable} on average).
For example, \cref{fig:no-transformations-possible-animation} shows
a code snippet of a project that could not be transformed:
The project contains six sprites, which in turn consist only of small
scripts (cf. \cref{fig:no-transformations-possible-animation}). Even though the scripts are not purely sequential animations, they provide no opportunity for transformation.

\begin{table}[tb]
  \centering
  \caption{Projects improved on combinations of objectives.}
  \label{table:improvements}
  \begin{tabular}{%
    l
    r%S[table-format=3,round-mode=none,round-integer-to-decimal=false]
    r%S[table-format=3,round-mode=none,round-integer-to-decimal=false]
    S[table-format=1.1]  @{}%
  } \toprule
    Objectives
        & \multicolumn{2}{c}{\# projects significantly}
                & \multicolumn{1}{c}{avg. \effectsize} \\
        & \multicolumn{1}{c}{decreased}
            & \multicolumn{1}{c}{increased}
                & \\ \midrule
    block size
        & \blocksImproved
            & \blocksWorsened
                & \blocksAvgEffect \\
    complexity
        & \complexityImproved
            & \complexityWorsened
                & \complexityAvgEffect \\
    entropy
        & \entropyImproved
            & \entropyWorsened
                & \entropyAvgEffect \\ \midrule
    block size and complexity
        & \blocksComplexityImproved
            & \blocksComplexityWorsened
                & \blocksComplexityAvgEffect \\
    block size and entropy
        & \blocksEntropyImproved
            & \blocksEntropyWorsened
                & \blocksEntropyAvgEffect \\
    complexity and entropy
        & \complexityEntropyImproved
            & \complexityEntropyWorsened
                & \complexityEntropyAvgEffect \\ \midrule
    block size, complexity, entropy
        & \threeObjectivesImproved
            & \threeObjectivesWorsened
                & \threeObjectivesAvgEffect \\ \bottomrule
  \end{tabular}
\end{table}

\Cref{table:improvements} summarises the results in terms of the
three objectives, considering the entire Pareto front. Out of
\num{\projectsRefactored} projects on which transformations were applicable, an
average of \num[round-precision=2]{\dominatingSolutions} projects were statistically significantly
improved with respect to all objectives, which is over
\percentageProjectsDominated of the projects on which
code transformations were applicable.
Note that \cref{table:improvements} checks for strict improvement in all listed
dimensions. However, a total of \projectsWithDominatingSolution~projects result in dominating solutions, i.e., solutions where no objective is worse and at least one is improved.

Overall, \num{\blocksImproved} projects were significantly improved regarding
their block size, \num{\complexityImproved} regarding their complexity, and
\num{\entropyImproved} regarding their entropy.
The lower number of improvements of size is not surprising, as most
transformations do not explicitly target size, and may in fact increase size.
For example, each split of a script introduces at least one additional block
for the additional event handler block. Indeed we included size as one of the
optimisation goals mainly to prevent size from \emph{growing} excessively,
rather than trying to reduce it.
Note that, even though the effect size of \num{\blocksAvgEffect} suggests that
size is more likely to increase than decrease on average, it may still be the
case that solutions with smaller size exist on a Pareto front, if the search
found no solutions that \emph{dominate} the original program with respect to
all objectives (RQ2 explicitly looks at improvements in each dimension).

The number of projects in which size was improved together with either
complexity or entropy is substantially smaller than the number of cases where
it was possible to improve these objectives individually
(\cref{table:improvements}).
However, in those cases where only complexity and entropy were improved, size
did not increase by a lot (cf. \cref{sec:evaluation-rq2}). Consequently, many
improved solutions were found that did not dominate the original program, but
still increase our measured objectives. Over
\percentageProjectsImprovedComplexityEntropy of all projects and over
\percentageRefactorableProjectsImprovedComplexityEntropy of all refactored
programs were improved for both of these objectives together.
Consequently, it might be worth exploring explicitly whether our anticipation
of size increase would manifest, or whether optimising only for complexity and
entropy achieves similar or better results. Alternatively, it might be
interesting to define additional transformations that explicitly aim to improve
size.
%

%Overall,
%we can answer our first research question as follows:
%
\summary{RQ1}{%
  \percentageRefactorableProjectsImprovedComplexityEntropy of the projects were
  improved in terms of complexity and entropy, and  
  \projectsWithDominatingSolution~projects resulted in dominating solutions.
%  \percentageProjectsDominated of transformable projects were significantly improved in \emph{all} metrics.
  }

%%%%%%%%%%%%%%%%%%%%%%%%%%%%%%%%%%%%%%%%%%%%%%%%%%%%%%%%%%%%%%%%%%%%%%%%%%%%%%%%
\subsection{RQ2: Changes in Complexity, Entropy and Size}\label{sec:evaluation-rq2}

\begin{figure}[t]
  \centering
  \begin{subfigure}[t]{0.29\linewidth}
      \centering
      \includegraphics[height=10.5em]{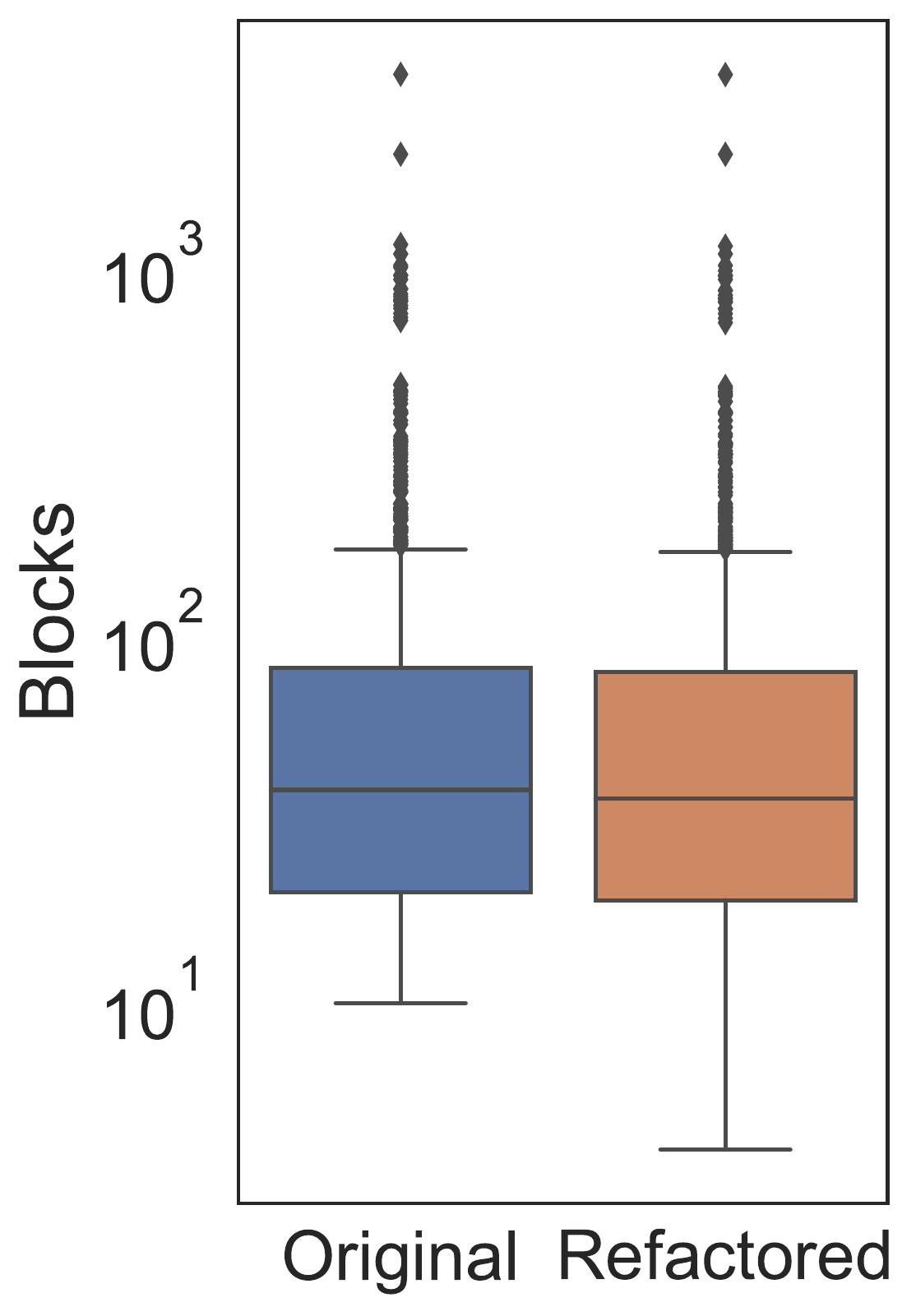}
      \caption{Log \#blocks}
      \label{fig:blocks_comparison}
  \end{subfigure}
  \hfill
  \begin{subfigure}[t]{0.32\linewidth}
      \centering
      \includegraphics[height=10.5em]{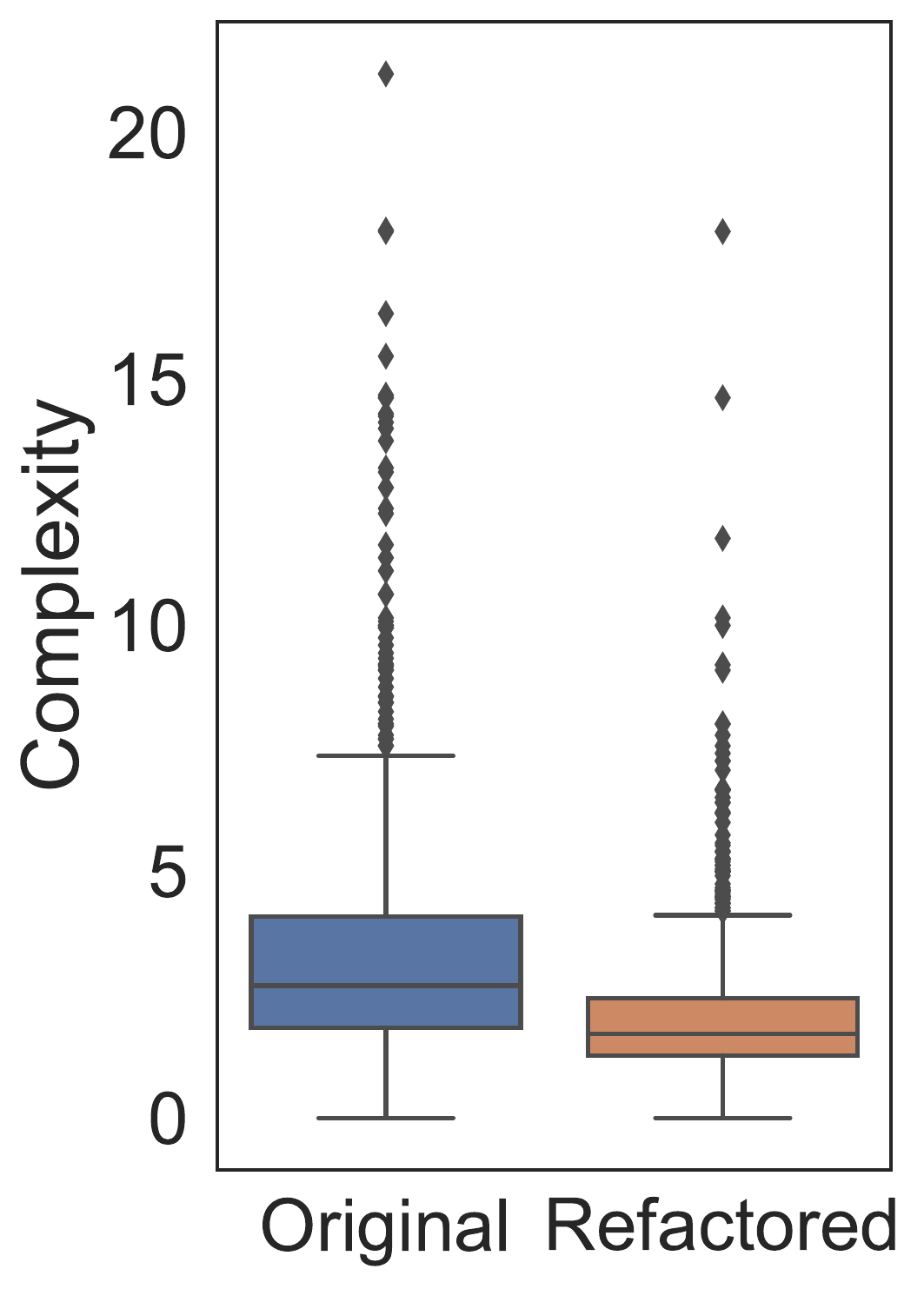}
      \caption{Halstead difficulty}
      \label{fig:complexity_comparison}
  \end{subfigure}
  \hfill
  \begin{subfigure}[t]{0.29\linewidth}
      \centering
      \includegraphics[height=10.5em]{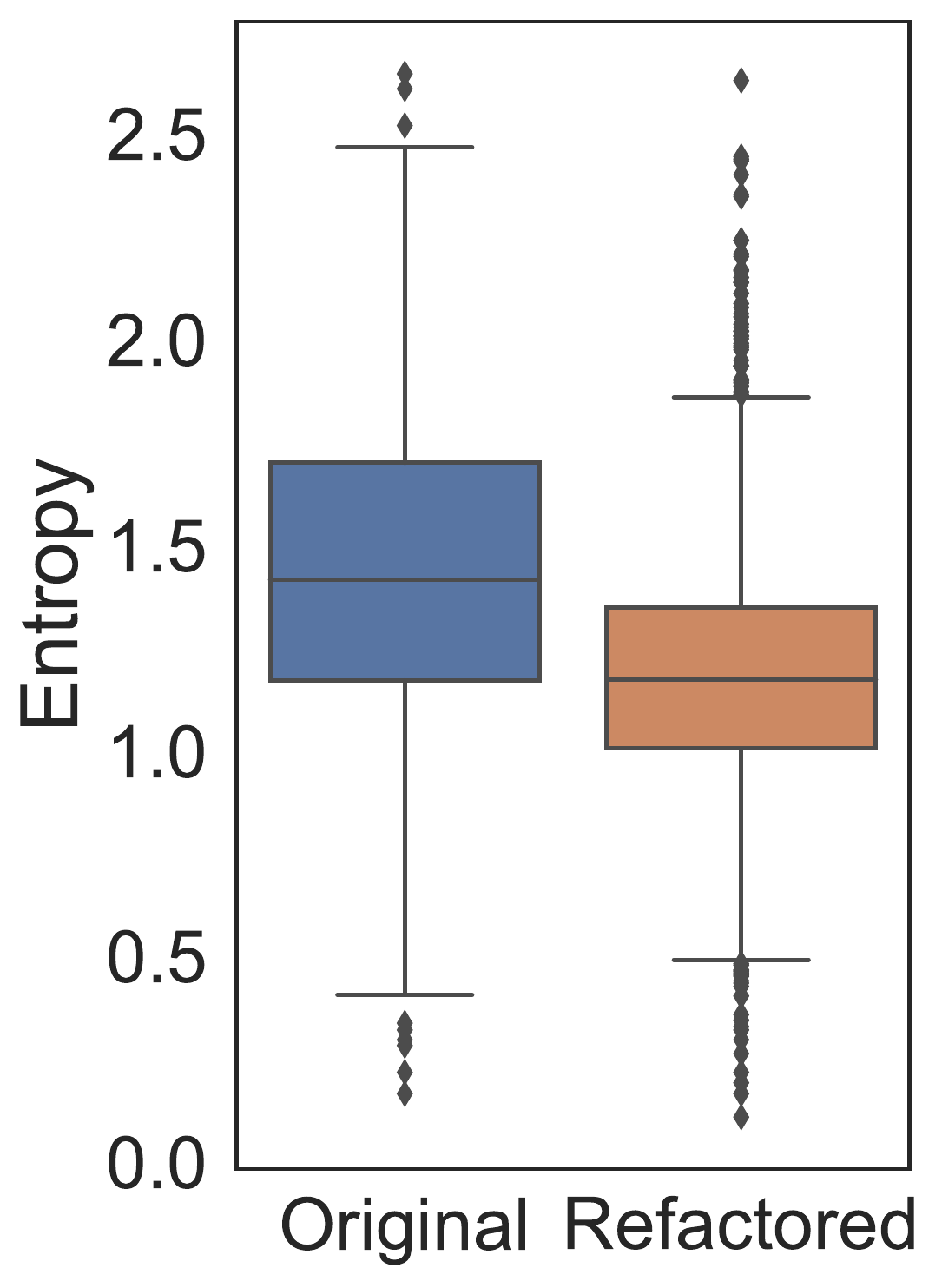}
      \caption{Category entropy}
      \label{fig:entropy_comparison}
  \end{subfigure}
  \caption{Comparison of the metrics before and after applying search-based refactorings. Plots show the best value on the Pareto front, averaged over 30 runs.}
\end{figure}

\Cref{fig:blocks_comparison} summarises the differences in terms of the number of blocks for each of the projects, \cref{fig:complexity_comparison} in terms of the complexity, and \cref{fig:entropy_comparison} in terms of the entropy.

\begin{figure}[t]
    \centering
    \begin{subfigure}[t]{0.49\linewidth}
      \centering
      \includegraphics[scale=0.16]{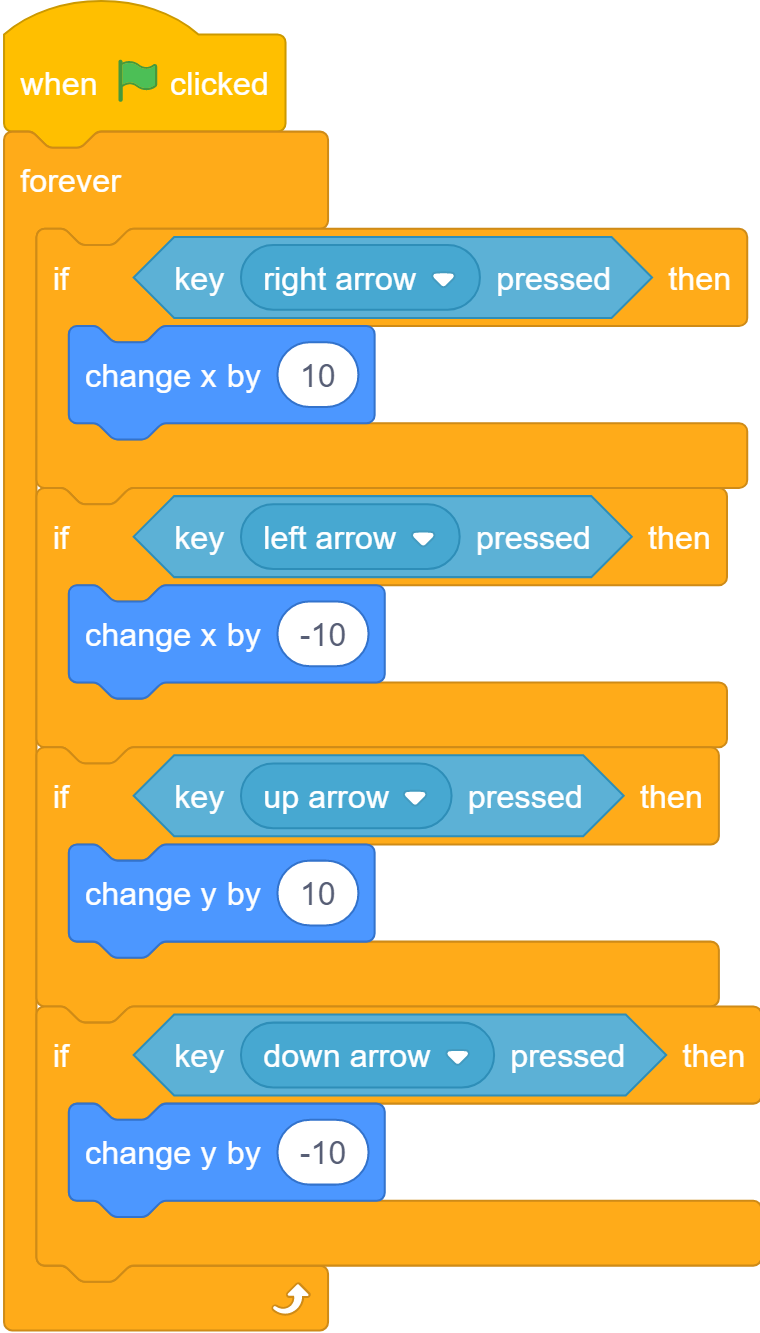}
      \caption{Original program}
      \label{fig:example_size}
    \end{subfigure}
    \begin{subfigure}[t]{0.49\linewidth}
      \centering
      \includegraphics[scale=0.16]{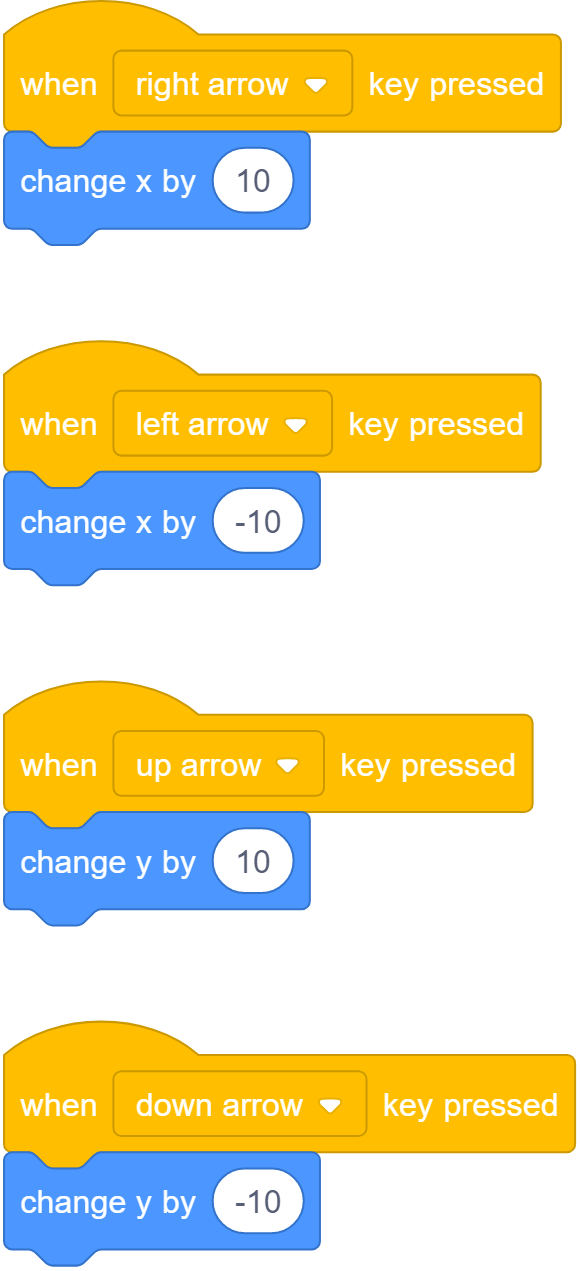}
      \caption{Refactored version}
      \label{fig:example_refactoredSize}
    \end{subfigure}
    \caption{Size of the script decreases significantly.}
    \label{fig:size_example}
\end{figure}

On average, projects have \num[round-precision=2]{\meanBlocksBeforeRefactoring}
blocks before the refactoring, and
\num[round-precision=2]{\meanBlocksAfterRefactoring} after, considering the
smallest projects on each Pareto front ($p \blocksAfterRefactoringP$).
\Cref{fig:size_example} shows an example where the size is reduced
successfully: A complex script is split into several smaller event handler
scripts.
Indeed our refactorings tend to significantly increase the number of scripts
($p \scriptsP$). This is an expected result, given that two of the most
frequent transformations split scripts into two or more scripts (cf.
\cref{sec:evaluation-rq3}). As a consequence of scripts being split, we find
that the number of long script code smells significantly decreases ($p
\longscriptsP$). In total, \num{\longscriptsDiff} long script smells were
removed on average per run over the \num{1000} projects.

The size objective appears to be harder to improve than the other two
objectives, which is due to the transformations we defined. Splitting scripts
may increase the number of blocks, since each additional script adds a new
event handler block. Similarly, splitting loops or conditional statements may
increase the block size, even when improving complexity and entropy.
However, the number of added blocks tends to be small. Considering the mean
size over all individuals in each Pareto front (which may include the original
project if no dominating solution was found), the block size increases from
\num[round-precision=2]{\meanBlocksBeforeRefactoringMean} to
\num[round-precision=2]{\meanBlocksAfterRefactoringMean} on average over all
projects and runs.
%For example,
%\cref{fig:non-drastic-size} shows a refactoring that decreases complexity and
%entropy while adding a few blocks.

\begin{figure}[t]
    \centering
    \begin{subfigure}[t]{.49\linewidth}
      \centering
      \includegraphics[scale=0.125]{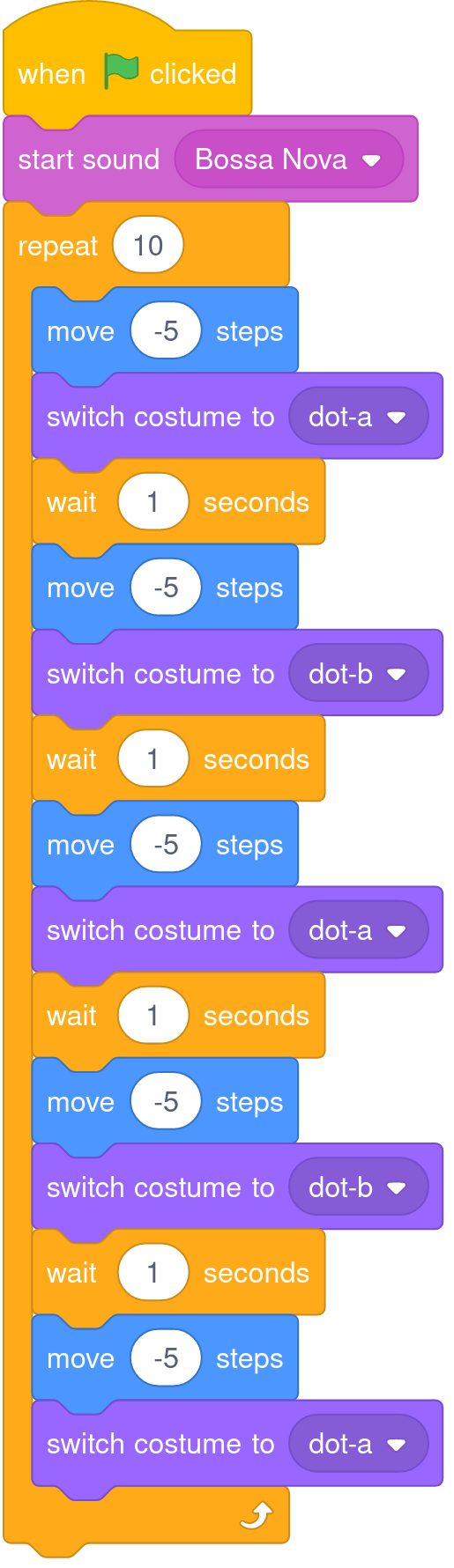}
      \caption{Original program}
      \label{fig:example_complexity}
    \end{subfigure}
    \begin{subfigure}[t]{.49\linewidth}
      \centering
      \includegraphics[scale=0.125]{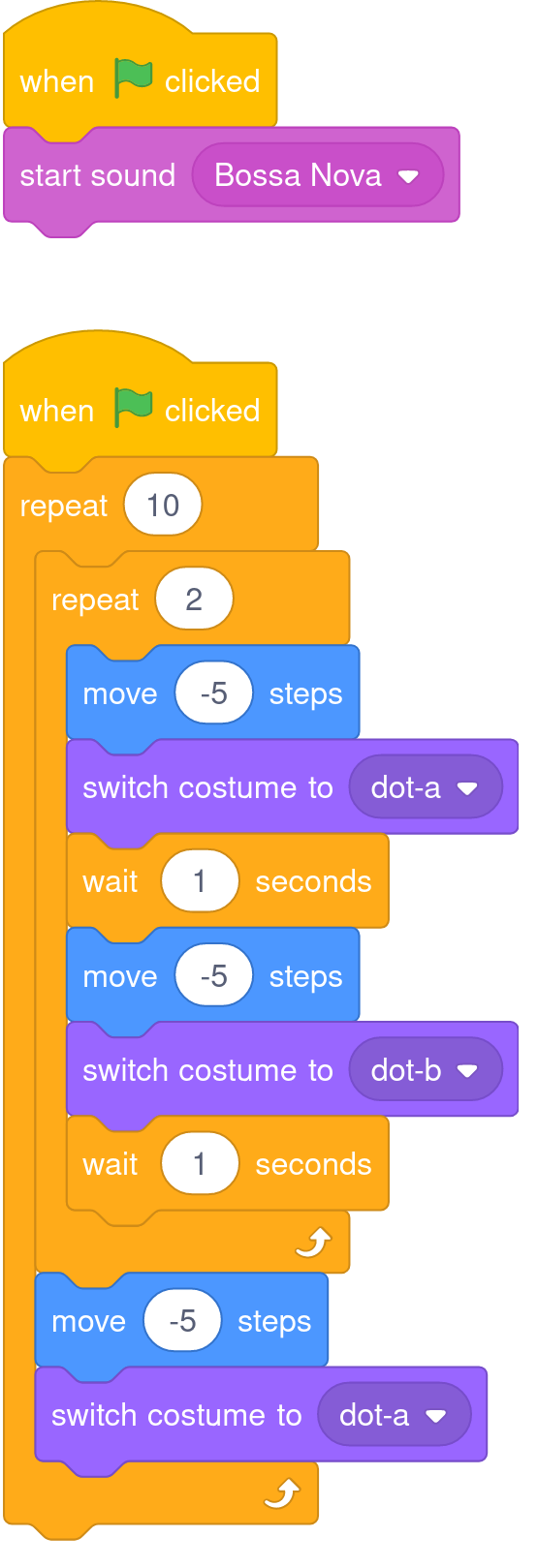}
      \caption{Refactored version}
      \label{fig:example_refactoredComplexity}
    \end{subfigure}
    \caption{Complexity of the script decreases from \num{8.0} to \num{1.58}.}
    \label{fig:complexity_example}
\end{figure}

The search succeeds in reducing the Halstead difficulty from an average of
\num[round-precision=2]{\meanComplexityBeforeRefactoring} to
\num[round-precision=2]{\meanComplexityAfterRefactoring} ($p
\complexityAfterRefactoringP$). \Cref{fig:complexity_example} is an example
where complexity is reduced: On the one hand, the replacement of the repetition
with a simple fixed number of iterations reduces the size of the script. On the
other hand, the extraction of the unrelated sound block leads to a very small
script, which likely skews the fitness calculation which is based on the
average per script. It might be worth experimenting with variants of the
fitness function (e.g., using the maximum or median) to further improve the
guidance for the search.

%
% \begin{figure}[t]
%   % Pictures taken from project id: 400000272, seed 0, frontid 1
%   \centering
%   \subfloat[\label{fig:before-non-drastic} Before refactoring]{
%     \centering
%     \includegraphics[width=0.45\columnwidth]{pictures/evaluation/complexityEntropyBefore}
%   }
%   %
%   \hfill
%   %
%   \subfloat[\label{fig:after-non-drastic} After refactoring]{
%     \includegraphics[width=0.45\columnwidth]{pictures/evaluation/complexityEntropyAfter}
%   }
%   %
%   \caption{Non drastic size change while improving complexity and entropy}
%   \label{fig:non-drastic-size}
% \end{figure}

%
% If we look more closely at the solutions that are generated, we see that the solutions that dominate have a significantly bigger project size. (Mann-Whitney U with
% $p < 0.05$).  So if we have more blocks, we can find more transformations to execute, with the result that we get dominating solutions. For non dominated solutions we get a mean of \num[round-precision=2]{\meannondominatedblocks} blocks and \num[round-precision=2]{\meandominatedblocks} blocks for dominated solutions.

%
%
%
% %
% For the size of the modified programs, we can say that the
% refactoring process lead in average to a decrease of blocks like it
% is shown in \cref{fig:blocks_comparison}. By applying the right code
% transformations, duplicates or unnecessary blocks can be removed
% or replaced by fewer blocks which reduces their number
% in a script which \cref{fig:size_example} demonstrates.
% %
% \todo[inline]{How often is size increased?}

%Our transformations The entropy and complexity objectives were easier
%to optimize on random \Scratch samples than on programs in other languages.

\begin{figure}[t]
    \centering
    \begin{subfigure}[t]{0.49\linewidth}
      \centering
      \includegraphics[scale=0.16]{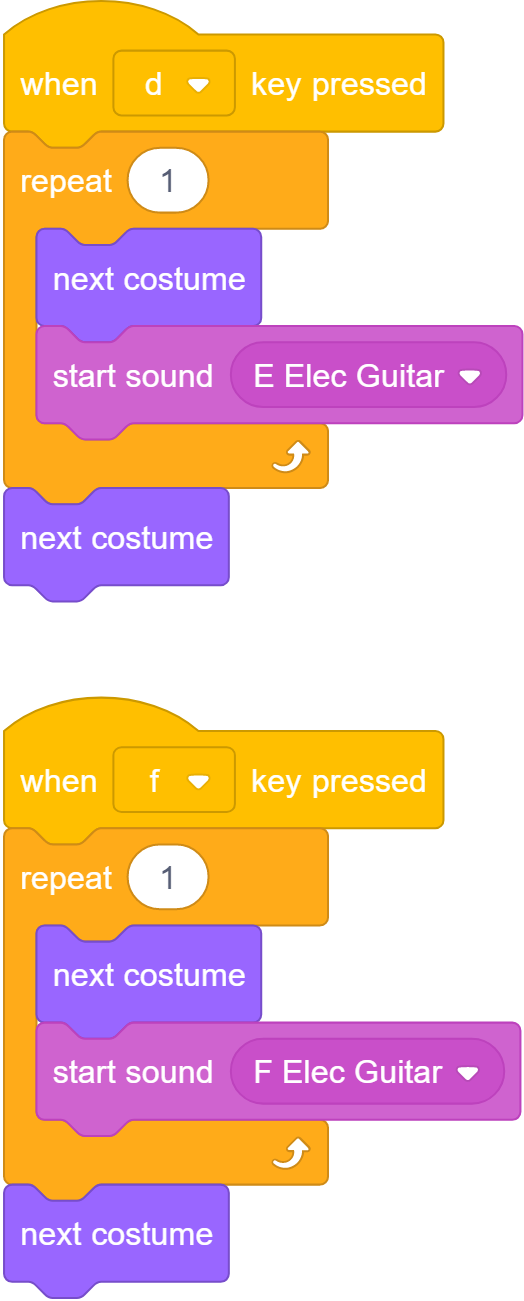}
      \caption{Original program}
      \label{fig:example_entropy}
    \end{subfigure}
    \begin{subfigure}[t]{0.49\linewidth}
      \centering
      \includegraphics[scale=0.16]{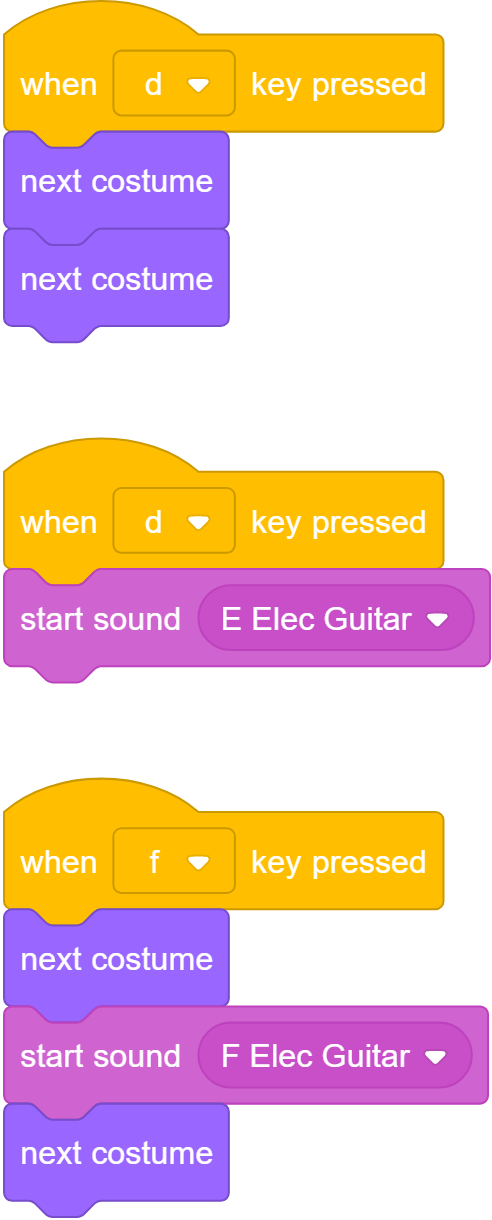}
      \caption{Refactored version}
      \label{fig:example_refactoredEntropy}
    \end{subfigure}
    \caption{Entropy of the script decreases from \num{1.75} to \num{1.12}.}
    \label{fig:entropy_example}
\end{figure}

\Cref{fig:entropy_comparison} shows that entropy is also successfully reduced
by the search: On average, the entropy decreases from
\num[round-precision=2]{\meanEntropyBeforeRefactoring} to
\num[round-precision=2]{\meanEntropyAfterRefactoring} ($p
\entropyAfterRefactoringP$).
The category entropy captures how chaotic (uncertain) scripts are with respect
to the categories of blocks they use. Intuitively, scripts that proportionally contain more blocks of a category have a lower entropy value.
\Cref{fig:entropy_example} shows an example where two scripts, which
have high entropy as they mix together different block categories, are
refactored to more coherent scripts. The entropy is improved because the first
script is split into two scripts that each have a distinct category of
blocks\footnote{It might have been the intention of the learner to have the
second costume change wait for completion of the sound, but since they did not
use a \scratchblocktable{play sound until done} block the transformation is
semantically equivalent.}, furthermore the useless \scratchblock{repeat (1)}
loops are ``unrolled'', thus removing all blocks of the control category.

Some of the transformations (e.g., \emph{Split Loop}) may increase the number of
control and operator blocks in a script, which the entropy fitness interprets
as an improvement. It might be worth investigating in the future whether the
category entropy should exclude control blocks.

%
%Overall, we answer our second research question as follows:
%
\summary{RQ2}{%
  Our approach tends to improve complexity and entropy by splitting scripts into multiple simpler versions, which may add a small number of blocks.
}

%%%%%%%%%%%%%%%%%%%%%%%%%%%%%%%%%%%%%%%%%%%%%%%%%%%%%%%%%%%%%%%%%%%%%%%%%%%%%%%%

\subsection{RQ3: Program Transformations}\label{sec:evaluation-rq3}

\begin{figure}
  \centering
  \includegraphics[width=\columnwidth]{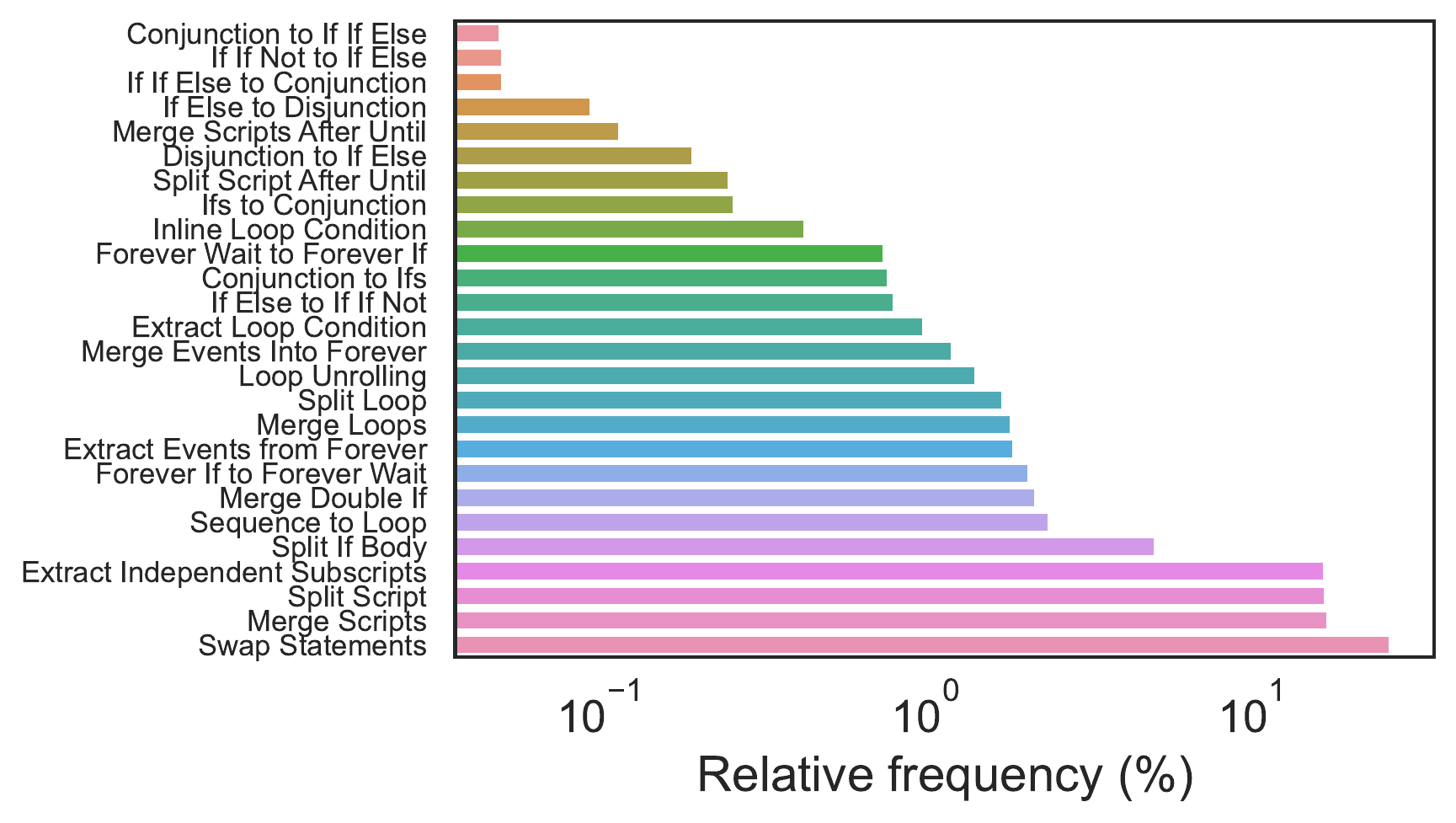}
  \caption{Frequency of transformations in the optimal solutions of \num[round-precision=2]{30} runs on \num[round-precision=4]{1000} programs (log scale).}
  \label{fig:applied-transformations}
\end{figure}

Across all solutions the mean number of transformations is
\num{\transformationsmean}. For each program we produced an average of
\num{\meanfrontsize} unique solutions.
\cref{fig:applied-transformations} shows the frequencies of the different
transformations relative to the \num[round-precision=6]{\transformationscount}
transformations aggregated across all runs.
The most frequent transformation is \emph{Swap Statements} with a relative
frequency of \swappercent, followed by \emph{Merge Scripts} with
\mergescriptspercent, \emph{Split Script} with \splitscriptpercent and
\emph{Extract Independent Subscripts} with \splitslicepercent.

\Cref{fig:rq3-example} shows a small example program which shows how the
frequent transformations form a refactoring which leads to a dominating solution.
The initial script consists of duplicate \scratchblock{if}
blocks which are separated by a \scratchblock{say} block. The most promising
transformation to apply here is the \emph{Merge Double If} transformation, which
merges the duplicate \scratchblock{if} blocks.
However, merging requires two successive \scratchblock{if} blocks. There are
several possibilities to reach the intermediary state in which merging is possible.
Either, apply a \emph{Swap Statements} transformation on one \scratchblock{if}
and the \scratchblock{say} block. Or, alternatively, split and merge the script in the suitable order,
i.e., apply \emph{Split Script} or \emph{Extract Independent Subscripts}
and merge the split parts back in the suitable order with \emph{Merge Scripts} to merge the double \scratchblock{if}s.

%
%This example illustrates the usefulness of the frequent refactorings, particularly,
%in combination. 
In general, splitting scripts can reduce the entropy of the initial script
which may lead to dominating solutions. Merging scripts can be an important
intermediate step for further simplifications, and it reduces the number of
blocks due to the hat block which is no longer duplicated. However, the final
result in \cref{fig:rq3-example} also illustrates an unnecessary application of
the \emph{Swap Statements} transformation: The \scratchblock{turn degrees} and
the \scratchblock{move steps} blocks have been swapped unnecessarily, and since
\emph{Swap Statements} is the only refactoring which is its own inverse, it
may be applied repeatedly. In general, it
might be useful to apply minimisation on refactoring sequences before
presenting solutions to the user in order to avoid redundant transformations,
and it might also be a viable alternative to minimise the size of the
refactoring sequence, rather than the program itself, as one of the search
objectives.

The least frequent transformations are \emph{If If Else to Conjunction}, which
was only applied \ififelsetoconjunctioncount times, and \emph{If If Not to If
Else} with \ififnottoifelsecount transformations, followed by \emph{Conjunction
to If If Else} with \conjunctiontoifelsecount occurrences.
There are multiple possible reasons for this. First, \emph{If If Else to
Conjunction} worsens all three metrics: Even though the control flow is
simplified, the total number of blocks and the complexity are increased since
all constituent operator blocks are counted. The same observation also applies
to the \emph{If If Not to If Else} transformation. 
Second, these transformations have quite strict preconditions on the combination of blocks to be applicable and other transformations may interfere.
However, these transformations may be important prerequisites to enable other
transformations by resolving nested control flow.
Besides these transformations, we note that splitting transformations seem to occur more often than merging transformations, and we conjecture that this is because splitting will in most cases improve entropy and complexity, even if the number of blocks is increased.

\begin{figure}[t]
  \centering
  \begin{subfigure}[t]{.49\linewidth}
    \centering
    \includegraphics[scale=0.18]{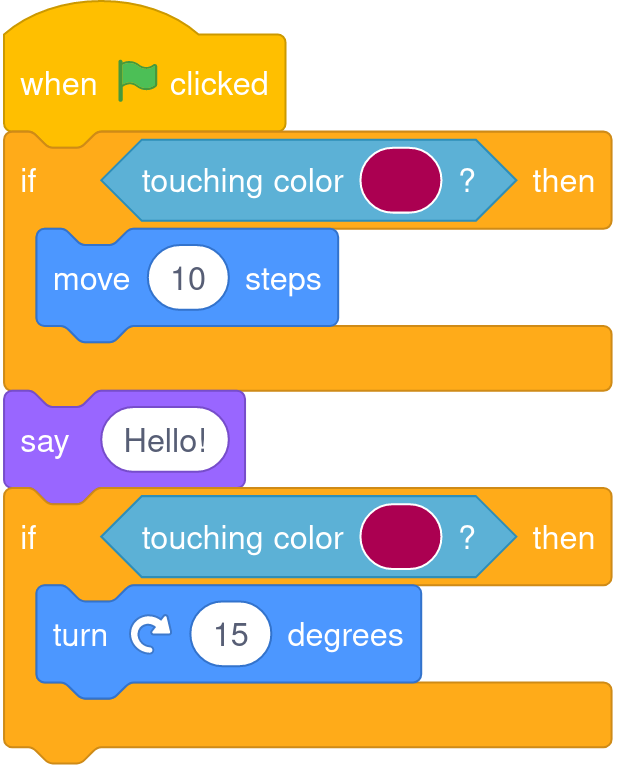}
    \caption{Original program}
  \end{subfigure}
  \begin{subfigure}[t]{.49\linewidth}
    \centering
    \includegraphics[scale=0.18]{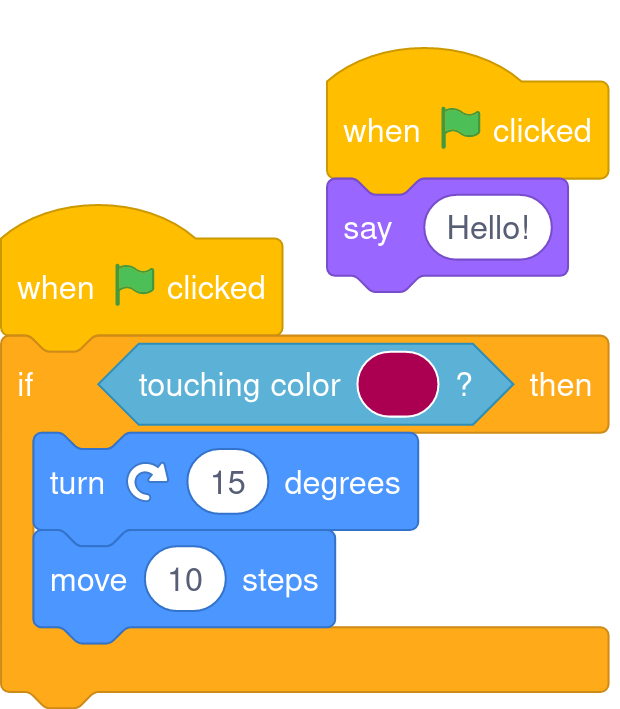}
    \caption{Refactored program}
  \end{subfigure}
  \caption{A script dominated by its refactored version.} % TODO better caption
  \label{fig:rq3-example}
\end{figure}

%
% \begin{figure}[t]
%   \centering
%   \subfloat[\label{fig:non-dominated-smaller}]{
%     \centering
%     \includegraphics[width=0.55\columnwidth]{pictures/evaluation/NonDominatedSmaller}
%  }
%   %
%   \hspace*{-.25cm}
%   %
%   \subfloat[\label{fig:pareto-fronts-size}]{
%     \includegraphics[width=0.55\columnwidth]{pictures/evaluation/SizeParetoFront}
%   }
%   %
%   \caption{Programs for which a dominating solution is found have more blocks. Generally,
%   there are multiple non-dominating solutions in the first front.}
% \end{figure}

%\todo[inline]{Fix labels of figures in Jupyter Notebook}
%\todo[inline]{Fix layout of figures (can one still read them when fixed or are they too small?)}
%\todo[inline]{Add meaningful captions where necessary}
%\todo[inline]{Ideally replace the example by an example taken from our dataset}
%\todo[inline]{Effect sizes?}
%\todo[inline]{Update all the numbers once we know which one is the final run}
%
%Overall,
%we can answer our third research question as follows:
%
\summary{RQ3}{%
  In our experiments, solutions are transformed \num{\transformationsmean} times on average, and the most common transformations are \emph{Swap Statements}, \emph{Merge Scripts} and \emph{Split Scripts}.
}

\section{Related Work}\label{sec:related-work}

The aim of search-based refactoring~\cite{OC03} is to improve code quality by
applying meta-heuristic search algorithms which select sequences of
refactorings. These refactorings are typically guided by design
metrics~\cite{MT04,MG18,MARIANI201714}. Many of the design metrics used to
measure aspects of object-oriented programs are not directly applicable to
\Scratch programs, and not all common refactorings are possible using the
abstraction mechanisms provided by the \Scratch language. For example, a
classical ``move method'' refactoring in \Scratch would only be possible for
scripts that only write global variables. 

An ``extract method'' refactoring could be partially performed using the
concept of custom blocks in \Scratch, but custom blocks cannot be re-used
across sprites. Techapalokul and Tilevich~\cite{TT19} nevertheless proposed
such a refactoring as part of their set of automated refactorings for \Scratch
programs. We intentionally avoided this type of refactoring since this would
introduce programming language aspects that early learners typically are not
accustomed with, and which might thus inhibit learning progress; our
refactorings do not add new language features that are not already contained in
a target program. Techapalokul and Tilevich~\cite{TT19} define specific refactorings for specific code smells, and apply all refactorings that are applicable exhaustively. In contrast, our approach applies search to explore the larger space of possible program variants independently of concrete preconditions or code smells.

%Furthermore, their refactorings require
%strict preconditions such as duplicate code or broad variable scopes.
%Therefore, if a program has no duplicate code or variables with broad scope, no
%refactoring is applied, even though it can still be unnecessarily complex and
%hard to read. In contrast to that, our atomic code modifications do not require
%such strict preconditions, but get powerful through suitable combination during
%search. They can transform the program to new intermediary states which enable
%beneficial modifications which would not have been possible on the initial program.

Since we transform code directly, the approach is reminiscent of Genetic
Improvement~\cite{petke2017genetic}: In principle our search can be seen as a
Genetic Algorithm that is applied using an existing program as starting point.
In contrast to Genetic Improvement, our modification operators are designed to
preserve the program semantics, since our optimisation goal is code quality,
rather than orthogonal aspects such as functionality or performance. Our
representation of refactoring sequences also differs %from common types 
in terms of the encoding used~\cite{petke2019survey} since
explainability of the transformations is an important aspect for our use case.

The number of code transformations that do not change the program semantics
found by our search is substantial, which provides further evidence for the
concept of neutral program
space~\cite{harrand2019journey,renzullo2018neutrality}. It is conceivable that,
given the target audience of novice programmers, the code we are attempting to
improve is less clean and contains more redundancy, thus providing even more
possibilities for semantics-preserving transformations.

\section{Conclusions}\label{sec:conclusions}

Although the block-based nature of introductory programming languages  makes it easy to assemble code, % in syntactically valid ways, 
it has
been observed that code quality is often problematic with young
learners~\cite{bugpatterns,hermans2016a,techapalokul2017understanding,meerbaum2011habits}, which may affect understanding and learning~\cite{HA16}.
We therefore introduced an automated approach to improve code quality for
\Scratch programs.

We envision different use cases for such an approach: First, teachers of young
programming learners very often are not skilled programmers themselves. If the
code they use to demonstrate programming to children contains problems, this
can be detrimental for the learning outcomes. Second, it is very common for
young programming learners to self-study using online tutorials and examples,
such that automated hints become important; this is also why we made
explainability a central component of our approach. Finally, learning support
systems (e.g.,~\cite{wang2020crescendo,marwan2020adaptive,price2017evaluation})
are often based on syntactically matching aspects of programs, and may fail in
the light of alternative solutions. An automated program transformation
approach may help to overcome this limitation.

While we demonstrated that search-based refactoring for \Scratch is feasible,
there is ample opportunity and need for further work. Although we applied
search techniques common for search-based refactoring, other algorithms and
parameters may lead to better solutions, and our set of code transformations
can likely be extended. Our implementation currently produces an entire Pareto
front of candidate solutions, but users might be overwhelmed, and so the
question of how to select one representative individual from the Pareto front
is an important one to answer. We used three metrics encoding important aspects
of code readability for our initial assessment, but alternative metrics may be
better suited to guide the search and to quantify desirable properties of the
target code. Furthermore, the explainability of generated refactorings needs
to be assessed. Ultimately, answering these questions will require qualitative
assessment and evaluation with teachers and children. To support future work,
\litterbox including our implementation of search-based refactoring is available at \url{https://github.com/se2p/LitterBox}.

\section*{Acknowledgements}
This work is supported by DFG project
FR 2955/3-1  ``Testing, Debugging, and Repairing
Blocks-based Programs''.

%
% ---- Bibliography ----
%
\bibliographystyle{IEEEtran}
\bibliography{related}

%
% ---- Appendix ----
%

\ifbool{inputappendix}{%
  % According to the IEEEtran documentation, it's a limitation of the LaTeX2e kernel that
% double column floats cannot be placed at the bottom of pages. To fix this, we need to
% load the stfloats pacakge, and use !b as float specifier for the first table.
% Furthermore, another limitation is that double column floats will never appear on the
% same page they were defined. Therefore, we have to input the first table BEFORE the
% \appendix command.

\begin{table*}[!b]
    \centering
    \caption{\label{table:concurrency-transformations}Atomic \Scratch code transformations based on the inherently concurrent nature of \Scratch programs.}
    \begin{tabular}{%
        @{}p{0.2\linewidth}p{0.2\linewidth} c p{0.35\linewidth}@{}%
    }
        \toprule
        Transformation & Initial/Transformed Script & Direction(s) & Transformed/Initial Script(s) \\ \midrule
        Split the body of a loop if there are no dependencies
        between the statements of its body and splitting does not create new
        ones.
        Replace the initial loop body by the first part of the body,
        add another loop for the remaining statements of the body.
        & \myincludegraphics[scale=0.15]{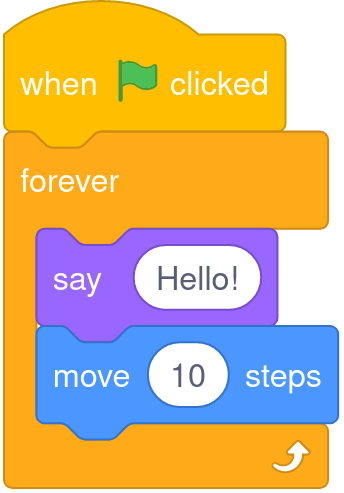}
        & \adjustbox{stack=cc}{\textbf{Split Loop} \\ $\rightleftarrows$ \\ \textbf{Merge Loops}}
        & \myincludegraphics[scale=0.15]{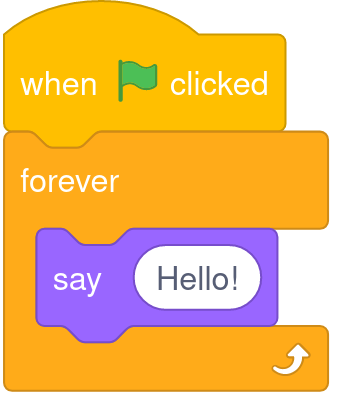}
        \myincludegraphics[scale=0.15]{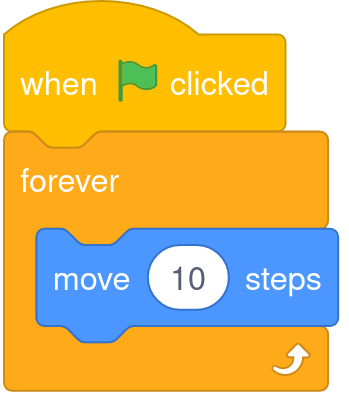} \\
        Split a script if there are no dependencies upwards.
        & \myincludegraphics[scale=0.15]{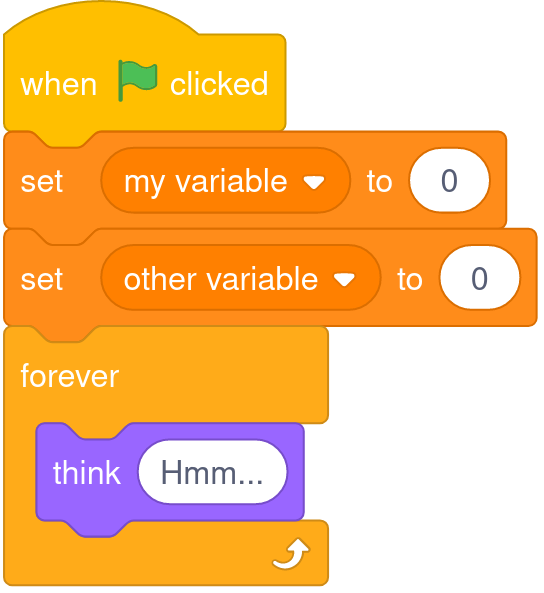}
        & \adjustbox{stack=cc}{\textbf{Split Script} \\ $\rightleftarrows$ \\ \textbf{Merge Scripts}}
        & \myincludegraphics[scale=0.15]{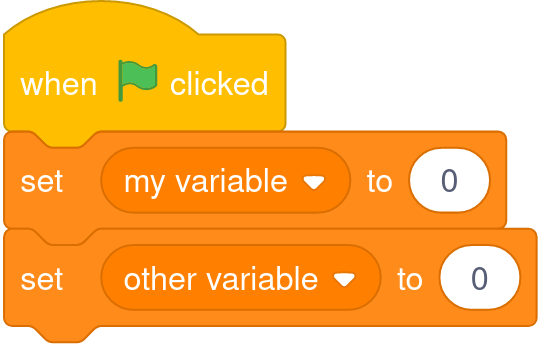}
        \myincludegraphics[scale=0.15]{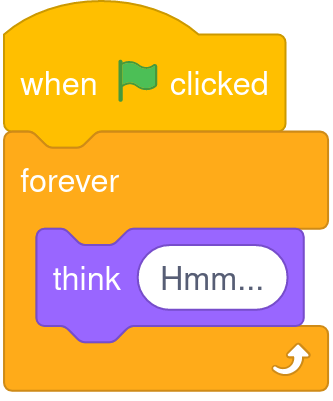} \\
        Split a script with dependencies into new scripts
        which do not depend on each other but respect the dependencies in the initial script.
        & \myincludegraphics[scale=0.15]{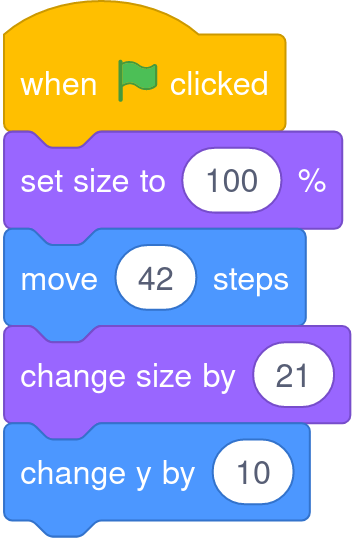}
        & \adjustbox{stack=cc}{\textbf{Extract Independent Subscripts} \\ $\rightarrow$ \\ ~}
        & \myincludegraphics[scale=0.15]{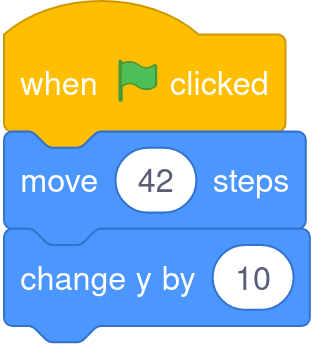}
        \myincludegraphics[scale=0.15]{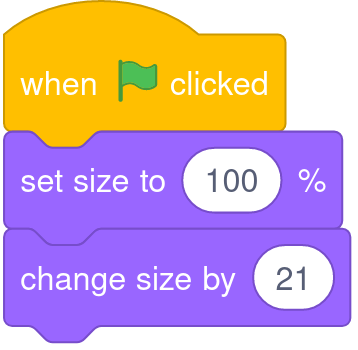} \\
        Replace a \scratchblocktable{forever} loop with \scratchblocktable{key pressed}
        conditionals by scripts triggered by \scratchblocktable{when key pressed} event listeners.
        & \myincludegraphics[scale=0.15]{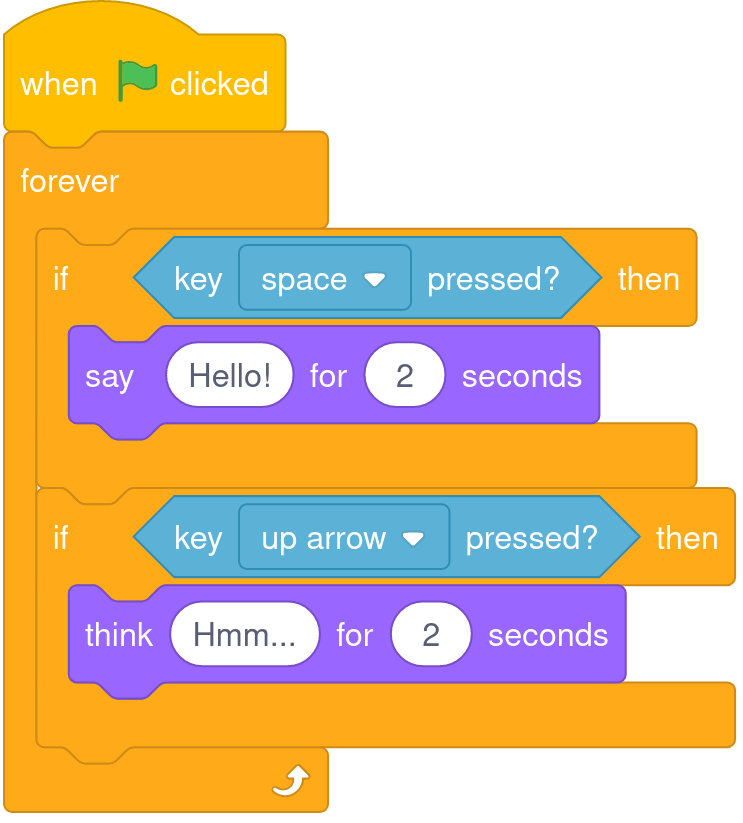}
        & \adjustbox{stack=cc}{\textbf{Extract Events from Forever} \\ $\rightleftarrows$ \\ \textbf{Merge Events into Forever}}
        & \myincludegraphics[scale=0.15]{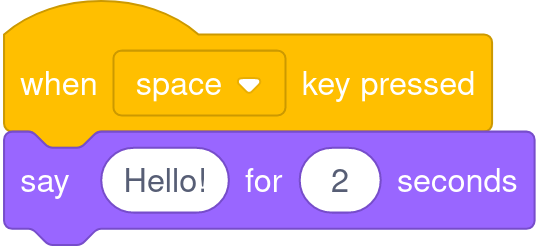}
        \myincludegraphics[scale=0.15]{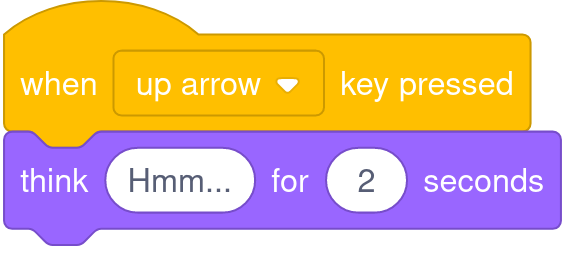} \\
        Split a script after a
        \scratchblocktable{repeat until} loop. Add a new script with a \scratchblocktable{wait until}
        and the same condition.
        & \myincludegraphics[scale=0.15]{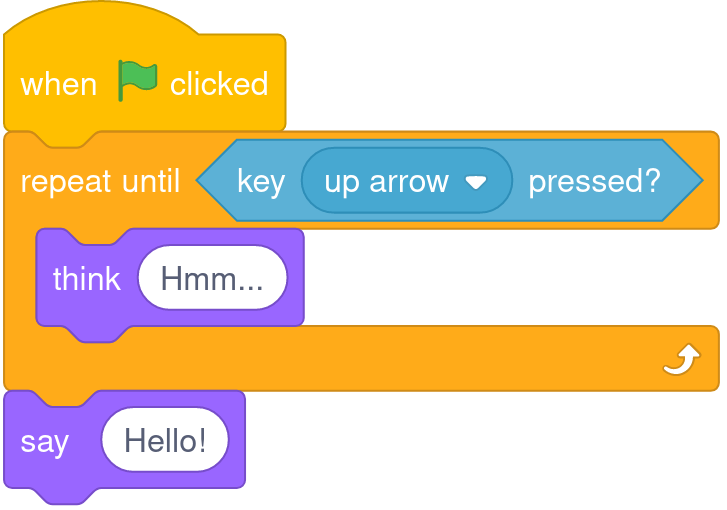}
        & \adjustbox{stack=cc}{\textbf{Split Script after Until} \\ $\rightleftarrows$ \\ \textbf{Merge Scripts after Until}}
        & \myincludegraphics[scale=0.15]{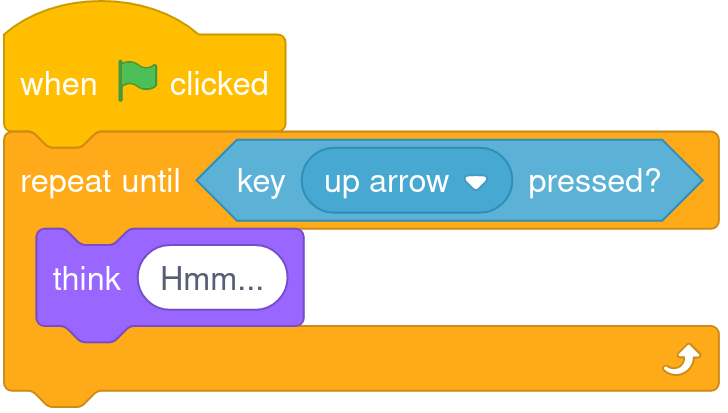}
        \myincludegraphics[scale=0.15]{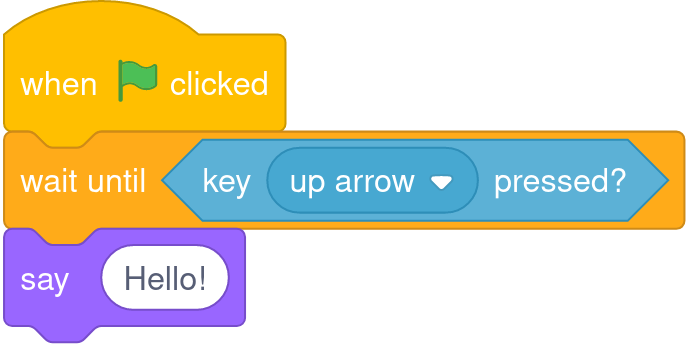} \\
        \bottomrule
    \end{tabular}
\end{table*}

\appendix

The following tables summarize the \Scratch code transformations implemented in our 
approach. 
\Cref{table:concurrency-transformations} showcases transformations enabled by the
inherently concurrent nature of \Scratch programs, allowing us to split a script into
several (smaller) scripts that are executed concurrently.
\Cref{table:control-transformations} illustrates transformations that
are driven solely by changing the way control flow is expressed in the given script, for
example, via loop unrolling or merging nested \scratchblock{if} blocks into a single
\scratchblock{if} block with multiple conditions.
We carefully designed all transformations to maintain program behaviour, taking into
account control, data, and time dependencies in the \Scratch code.
Most transformations can be applied in both directions (\( \rightleftarrows \)), while
others are uni-directional (\( \rightarrow \)).

\begin{table*}[htbp]
    \centering
    \caption{\label{table:control-transformations}Atomic control flow transformations for \Scratch programs.}
    \begin{tabular}{%
        @{}p{0.2\linewidth}p{0.2\linewidth} c p{0.4\linewidth}@{}%
    }
        \toprule
        Description & Initial/Transformed Script & Transformation/Inverse & Transformed/Initial Script\\ \midrule
        Swap two statements that are independent of each other, if swapping does not create new dependencies.
        & \myincludegraphics[scale=0.15]{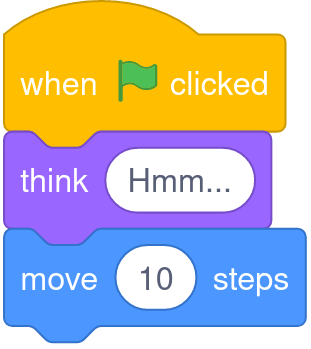}
        & \adjustbox{stack=cc}{\textbf{Swap Statements} \\ $\rightleftarrows$ \\ \textbf{Swap Statements}}
        & \myincludegraphics[scale=0.15]{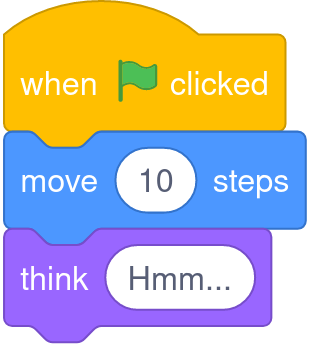}
        \\
        Replace a \scratchblocktable{repeat times} loop
        by the corresponding number of repetitions of its body.
        & \myincludegraphics[scale=0.15]{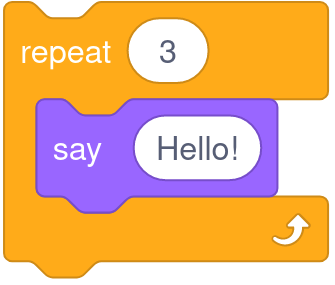}
        & \adjustbox{stack=cc}{\textbf{Loop Unrolling} \\ $\rightleftarrows$ \\ \textbf{Sequence to Loop}}
        & \myincludegraphics[scale=0.15]{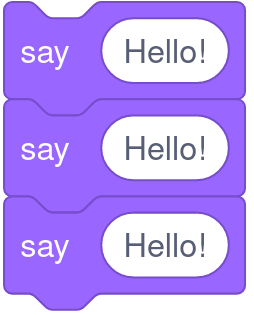}
        \\
        Replace an \scratchblocktable{if} block inside a \scratchblocktable{forever}
        loop with a \scratchblocktable{wait until} block with the same condition.
        & \myincludegraphics[scale=0.15]{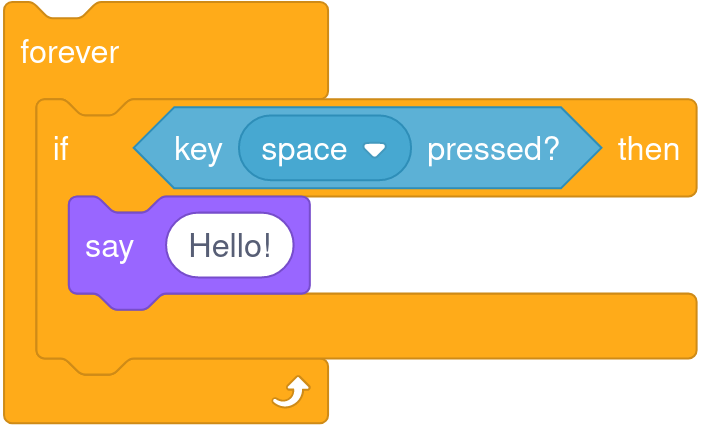}
        & \adjustbox{stack=cc}{\textbf{Forever If to Forever Wait} \\ $\rightleftarrows$ \\ \textbf{Forever Wait to Forever If}}
        & \myincludegraphics[scale=0.15]{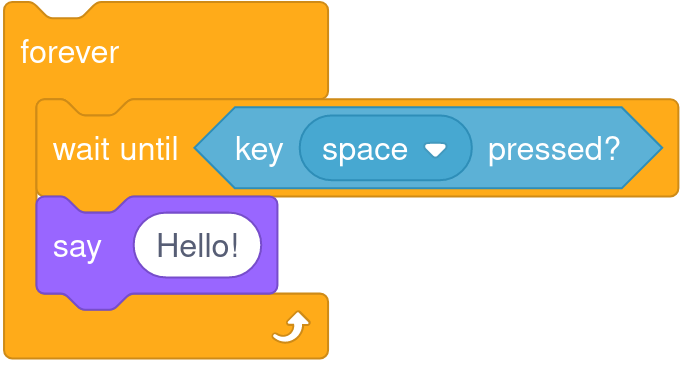}
        \\
        Transform a \scratchblocktable{forever} loop that
        conditionally terminates the script or the program to a \scratchblocktable{repeat until} loop.
        & \myincludegraphics[scale=0.15]{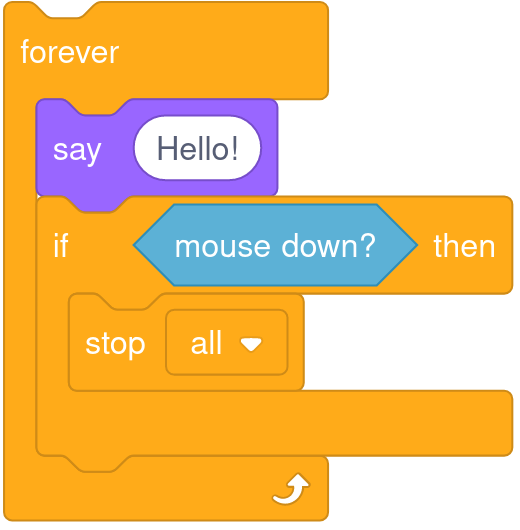}
        & \adjustbox{stack=cc}{\textbf{Extract Loop Condition} \\ $\rightleftarrows$ \\ \textbf{Inline Loop Condition}}
        &
        \myincludegraphics[scale=0.15]{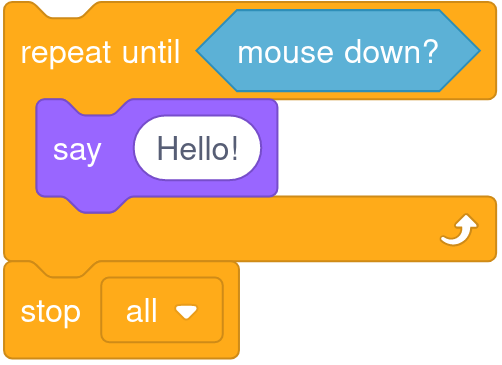}
        \\
        Split the body of an \scratchblocktable{if} block.
        Replace the \scratchblocktable{if} block by one containing the first part of the body,
        add another \scratchblocktable{if} for the remaining statements of the body.
        & \myincludegraphics[scale=0.15]{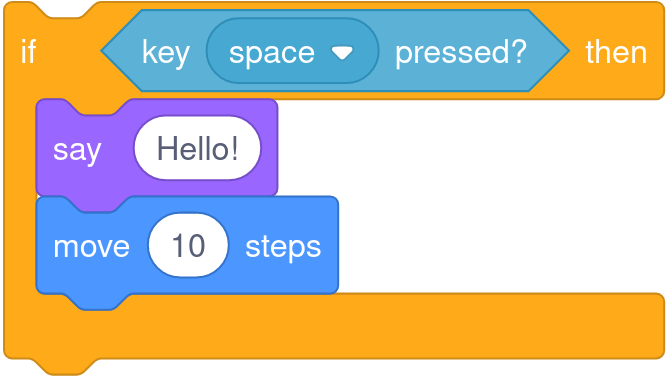}
        & \adjustbox{stack=cc}{\textbf{Split If Body} \\ $\rightleftarrows$ \\ \textbf{Merge Double If}}
        & \myincludegraphics[scale=0.15]{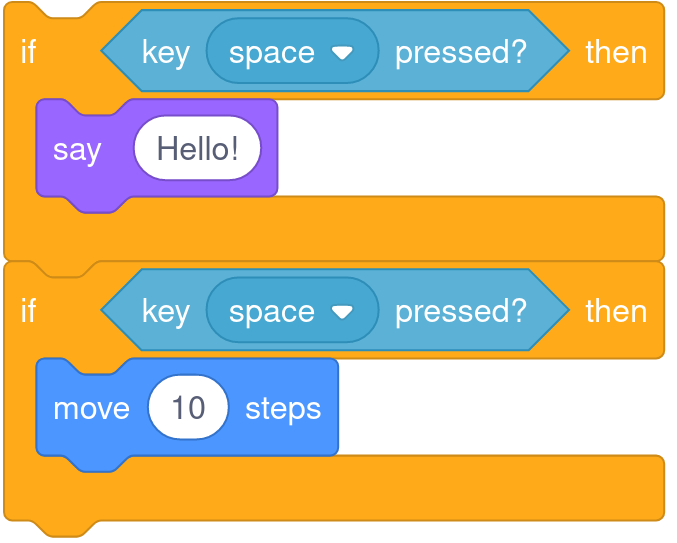}
        \\
        Split an \scratchblocktable{if else} block into two \scratchblocktable{if} blocks.
        The second \scratchblocktable{if} block checks on the negated initial condition.
        & \myincludegraphics[scale=0.15]{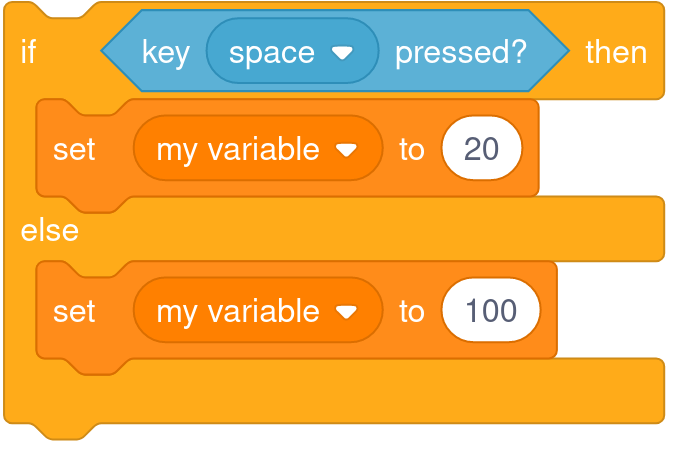}
        & \adjustbox{stack=cc}{\textbf{If Else to If If Not} \\ $\rightleftarrows$ \\ \textbf{If If Not to If Else}}
        & \myincludegraphics[scale=0.15]{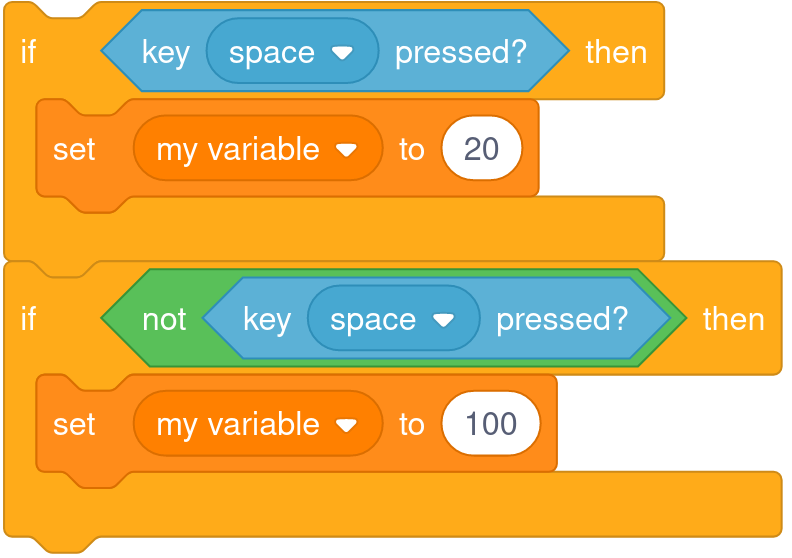}
        \\
        Transform two nested \scratchblocktable{if} blocks
        into an \scratchblocktable{if} block which checks for the conjunction of the initial conditions.
        & \myincludegraphics[scale=0.15]{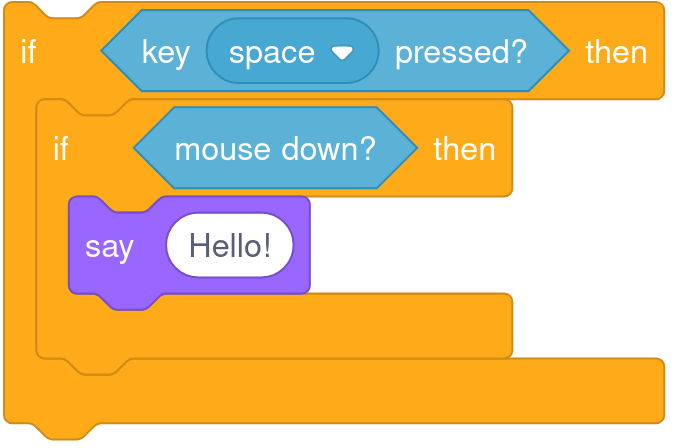}
        & \adjustbox{stack=cc}{\textbf{Ifs to Conjunction} \\ $\rightleftarrows$ \\ \textbf{Conjunction to Ifs}}
        & \myincludegraphics[scale=0.15]{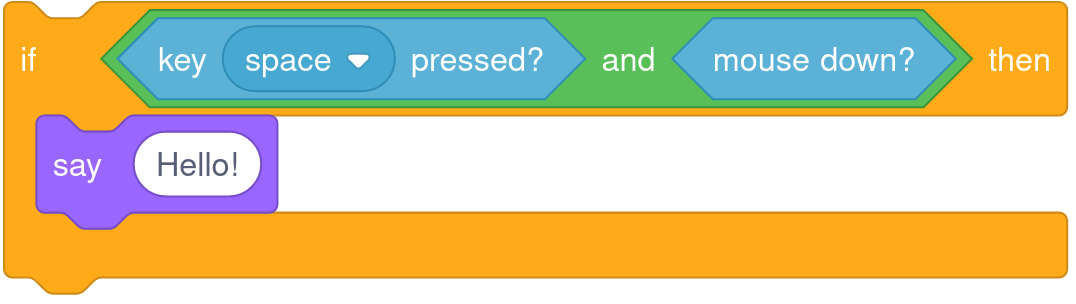}
        \\
        Replace an \scratchblocktable{if} containing an \scratchblocktable{if else} by two \scratchblocktable{if} blocks.
        The condition of the first \scratchblocktable{if} is the conjunction of the
        two initial conditions. The second condition is the condition of the initial first if block.
        & \myincludegraphics[scale=0.15]{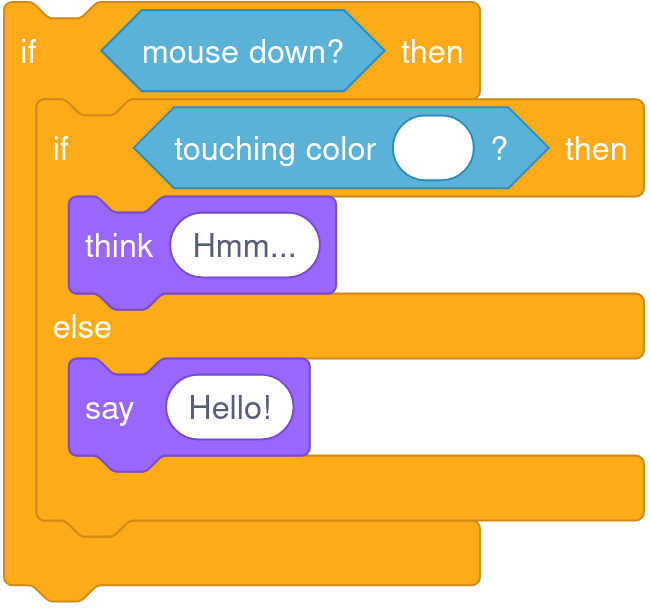}
        & \adjustbox{stack=cc}{\textbf{If If Else to Conjunction} \\ $\rightleftarrows$ \\ \textbf{Conjunction to If If Else}}
        & \myincludegraphics[scale=0.15]{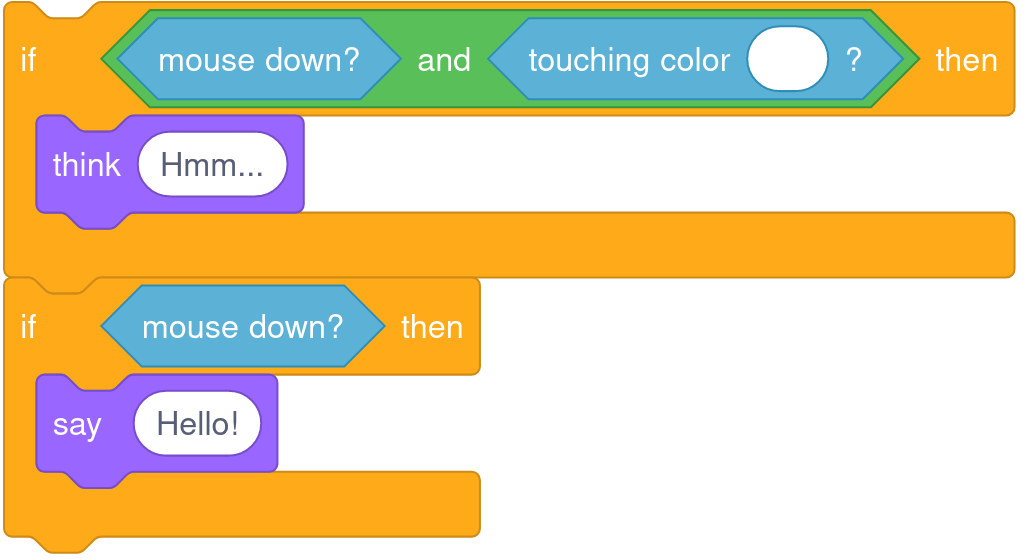}
        \\
        Replace an \scratchblocktable{if} block in the else case of an \scratchblocktable{if else}
        block by an \scratchblocktable{if} with the disjunction of the two conditions
        if the then cases of the initial conditionals have the same statements.
        & \myincludegraphics[scale=0.15]{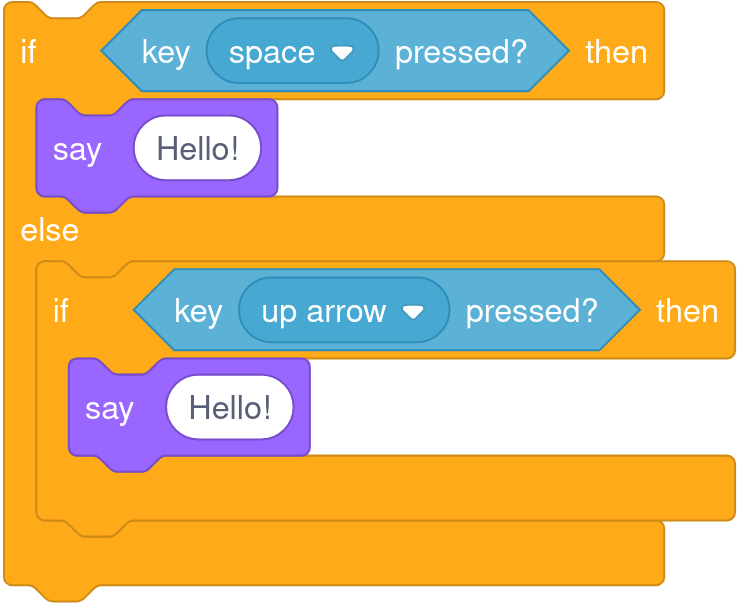}
        & \adjustbox{stack=cc}{\textbf{If Else to Disjunction} \\ $\rightleftarrows$ \\ \textbf{Disjunction to If Else}}
        & \myincludegraphics[scale=0.15]{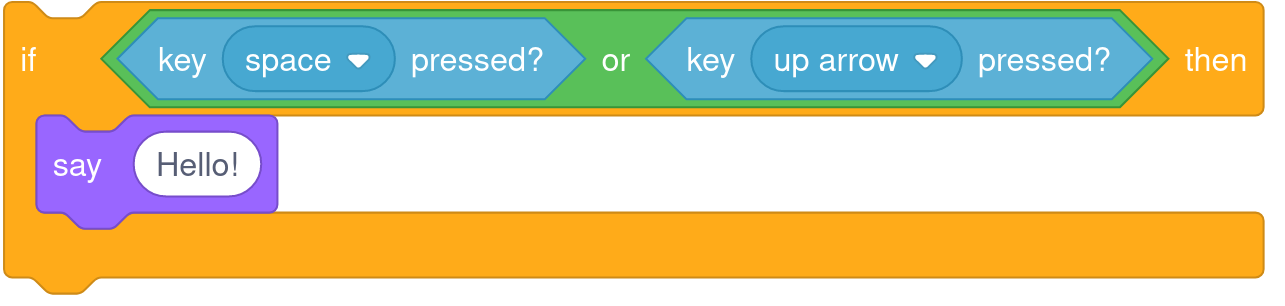}
        \\
        %
        % This is not implemented but might get implemented, keep it as a comment.
        % \textbf{Flatten nested ifs} Transform an if else block into two consecutive ifs.
        % & \myincludegraphics[scale=0.15]{pictures/transformations/flatten_nested_ifs_a} &
        % \myincludegraphics[scale=0.15]{pictures/transformations/flatten_nested_ifs_b} \\
        %
        \bottomrule
    \end{tabular}
\end{table*}

}{}

\end{document}